\documentclass[12pt]{article}

\begin{document}
\title{\Large \bf  Causal and Topological Aspects in Special and General Theory of Relativity}
\date{}

\maketitle\begin{center}{R \ V \ Saraykar$\ ^{*}$ and  Sujatha \ Janardhan$\ ^{\dag}$}  \\$\ ^{*}$ Department of Mathematics, R\ T\ M Nagpur University, Nagpur-440033. \\ e-mail :  ravindra.saraykar@gmail.com
\\ $\ ^{\dag}$ Department of Mathematics, St.Francis De Sales College,  Nagpur-440006. \\ e-mail : sujata\_jana@yahoo.com
\end{center}

\vspace{4mm}

\noindent In this article we present a review of a geometric and algebraic approach to causal cones and describe cone preserving transformations and their relationship with the causal structure related to special and general relativity. We describe Lie groups, especially matrix Lie groups, homogeneous and symmetric spaces and causal cones and certain implications of these concepts in special and general relativity, related to causal structure and topology of space-time. We compare and contrast the results on causal relations with those in the literature for general space-times and compare these relations with K-causal maps. We also describe causal orientations and their implications for space-time topology and discuss some more topologies on space-time which arise as an application of domain theory. For the sake of completeness, we reproduce proofs of certain theorems which we proved in our earlier work.
\vspace{2mm}

\noindent \emph{Key words :}  Causal cones, cone preserving transformations, causal maps, K-causal maps, space-time topologies, homogeneous spaces, causal orientations, domain theory.
\vspace{2mm}

\noindent Mathematics Subject Classification 2010 : \ 03B70, 06B35, 18B35, 22XX, 53C50, 54H15, 57N13, 57S25, 83A05, 83C05

\section{Introduction}

The notion of causal order is a basic concept in physics and in
the theory of relativity in particular. A space-time metric
determines causal order and causal cone structure. Alexandrov
[1,2,3] proved that a causal order can determine a
topology of space-time called Alexandrov topology which, as is now
well known, coincides with manifold topology if the space time is
strongly causal. The books by Hawking-Ellis, Wald and Joshi
[4,5,6] give a detailed treatment of causal structure of
space-time. However, while general relativity employs a Lorentzian
metric, all genuine approaches to quantum gravity are free of
space-time metric. Hence the question arises whether there exists
a structure which gets some features of causal cones (light cones)
in a purely topological or order-theoretic manner. Motivated by
the requirement on suitable structures for a theory of quantum
gravity, new notions of causal structures and cone structures were
deployed on a space-time.\\
The order theoretic structures, namely causal sets have been
extensively used by Sorkin and his co-workers in developing a new
approach to quantum gravity [7]. As a part of this program,
Sorkin and Woolgar [8] introduced a relation called
K - causality and proved interesting results by making use of
Vietoris topology. Based on this work and other recent work, S.
Janardhan and R.V.Saraykar [9,10] and E.Minguzzi [11,12]
proved many interesting results. Especially after good deal of
effort, Minguzzi [12] proved that K - causality condition is
equivalent to \emph{stably causal} condition.

\noindent More recently, K.Martin and Panangaden [13] making use of domain
theory, a branch of theoretical computer science, proved
fascinating results in the causal structure theory of space-time.
The remarkable fact about their work is that only \emph{order} is
needed to develop the theory and topology is an outcome of the
\emph{order}. In addition to this consequence, there are abstract
approaches, algebraic as well as geometric to the theory of cones
and cone preserving mappings. Use of quasi-order (a relation which
is reflexive and transitive) and partial order is made in defining
the cone structure. Such structures and partial orderings are used
in the optimization problems [14], game theory and decision making
etc [15]. The interplay between ideas from theoretical computer
science and causal structure of space-time is becoming more
evident in the recent works [16,17].

Keeping these developments in view , in this article, we
present a review of geometric and algebraic approach to
causal cones and describe cone preserving transformations and
their relationship with causal structure. We also describe certain
implications of these concepts in special and general theory of
relativity related to causal structure and topology of space-time.

Thus in section 2, we begin with describing Lie groups, especially
matrix Lie groups, homogeneous spaces and then causal cones. We
give an algebraic description of cones by using quasi-order.
Furthermore, we describe cone preserving transformations. These
maps are generalizations of causal maps related to causal
structure of space-time which we shall describe in section 3. We
then describe explicitly Minkowski space as an illustration of
these concepts and note that some of the space-time models in
general theory of relativity can be described as homogeneous
spaces.

In section 3, we describe causal structure of space time,
causality conditions, K-causality and hierarchy among these conditions
in the light of recent work of S. Janardhan and R.V.Saraykar and
E.Minguzzi and M.Sanchez [9,10,12,18]. We also describe
geometric structure of causal group, a group of transformations
preserving causal structures or a group of causal maps on a space-time.

\noindent In section 4, we describe causal orientations and their
implications for space-time topology. We find a parallel between
these concepts and concepts developed by Martin and Panangaden [13] to describe topology of space-time, especially a globally
hyperbolic one. Finally we discuss some more topologies on space-time which arise as an application of domain theory. Some material from Sections 2 and 4 is borrowed from the book by Hilgert and Olafsson [19].

We end the article with concluding remarks where we discuss more topologies which are different from, but physically more significant than manifold topology.


\section{ Lie Groups, Homogeneous Spaces, Causal Cones and cone preserving transformations}
\subsection{\emph{Lie Groups, Matrix groups and Homogeneous Spaces}}

To begin with, we describe Lie groups, matrix Lie groups,
homogeneous and symmetric spaces and state some results about
them. These will be used in the discussion on causal cones. We
refer to the books  [20,21] for more details.

\noindent\textbf{Definition 2.1.1:} \ Lie groups and matrix Lie groups:

\noindent \textbf{Lie group}: A finite dimensional manifold $\ G$ is
called a Lie group if $\ G$ is a group such that the group operations,
composition and inverse are compatible with the differential
structure on $\ G$. This means that the mappings\\
$\ G \times G  \rightarrow G : (x,y) \mapsto  x.y $ and \\
$\ G \rightarrow G : x \mapsto x^{-1} $ \\
are $\ C^{\infty}$ as mappings from one manifold to other.

\noindent The n-dimensional real Euclidean space $\ R^{n}$, n-dimensional
complex Euclidean space $\ C^{n}$, unit sphere $\ S^{1}$ in $\
R^{2}$, the set of all $\ n \times n $ real matrices $\ M(n,R)$
and the set of all $\ n \times n $ complex matrices $\ M(n,C)$ are
the simplest examples of Lie groups. $\ M(n,R)$ (and $\  M(n,C) $)
have subsets which are Lie groups in their own right. These Lie
groups are called \emph{matrix Lie groups}. They are important
because most of the Lie groups appearing in physical sciences such
as classical and quantum mechanics,  theory of relativity -
special and general, particle physics etc are matrix Lie groups.
We describe some of them here, which will be used later in this
article.\\
{$\ \textbf{Gl(n,R)}$} : General linear group of $\ n \times n $
real invertible matrices. It is a Lie group and topologically an
open subset of $\ M(n,R)$. Its dimension is $\ n^{2} $.\\
$\ \textbf{Sl(n,R)}$ : Special linear group of $\ n \times n $
real invertible matrices with determinant +1. It is a closed
subgroup of $\ Gl(n,R)$ and a Lie group in its own right, with
dimension $\ n^{2}-1 $.\\
$\ \textbf{O(n)}$: Group of all $\ n \times n $ real orthogonal
matrices. It is called an orthogonal group. It is a Lie group of
dimension $\ \frac{n(n-1)}{2}$.\\
$\ \textbf{SO(n)}$: Special orthogonal group- It is a connected
component of $\ O(n)$ containing the identity I and also a closed
(compact) subgroup of $\ O(n) $ consisting of real orthogonal
matrices with determinant +1. In particular $\ SO(2)$  is
isomorphic to $\ S^{1} $.\\
The corresponding Lie groups which are subsets of $\  M(n,C)$  are
$\ GL(n,C)$, $\ SL(n,C)$, $\ U(n)$ and $\ SU(n)$ respectively,
where \emph{orthogonal} is replaced by \emph{unitary}. $\ SU(n)$
is a compact subgroup of $\ GL(n,C)$. For n=2, it can be proved
that $\ SU(2)$ is  isomorphic to $\ S^{3}$, the unit sphere in $\
R^{4}$. Thus $\ S^{3}$ is a Lie group.[However for topological
reasons, $\ S^{2}$ is not a Lie group, though it is
$\ C^{\infty}$ -differentiable manifold]\\
$\ \textbf{O(p,q) and  SO(p,q)}$ : Let p and q be positive
integers such that $\ p+q = n$. Consider the quadratic form $\
Q(x_{1},x_{2}...x_{n})$ given by\\
$\ Q= x_{1}^{2} + x_{2}^{2}+... x_{p}^{2}-x_{p+1}^{2}-x_{p+2}^{2}...-x_{n}^{2}.$\\
The set of all $\ n \times n$  real matrices which preserve this
quadratic form Q is denoted by $\ O(p,q)$ and a subset of $\
O(p,q)$ consisting of those matrices of $\ O(p,q)$ whose
determinant is +1, is denoted by $\ SO(p,q)$. Both $\ O(p,q)$ and
$\ SO(p,q)$ are Lie groups. Here preserving quadratic form Q means
the
following:\\
Consider standard inner product $\ \eta$ on $\ R^{p+q} = R^{n}
$given by the diagonal matrix: \\
$\ \eta = diag(1,1 \ldots 1,-1,-1 \ldots -1)$, (1 appearing p times).\\
Then $\ \eta$ gives the above quadratic form $\
Q(x_{1},x_{2},...,x_{n})$,\\
i.e.$\  X \eta X^{T} = Q(x_{1},x_{2},...,x_{n})$ where $\ X=
[x_{1},x_{2},...,x_{n}]$. $\ n \times n$ matrix A is said to
preserve the quadratic form Q if $\ A^{T} \eta A = \eta $.

\noindent $\ O(p,q)$ is called \emph{indefinite orthogonal group} and $\
SO(p,q)$ is called \emph{indefinite special orthogonal group}.
Dimension of $\ O(p,q)$ is $\ \frac{n(n-1)}{2}$.\\
Assuming both p and q are nonzero, neither of the groups $\
O(p,q)$ or $\ SO(p,q)$ are connected. They have respectively four
and two connected components. The identity component of $\ O(p,q)$
is denoted by $\ SO_{o}(p,q)$ and can be identified with the set
of elements in $\ SO(p,q)$ which preserves both orientations.

\noindent In particular $\ O(1,3)$ is the Lorentz group, the group of all
Lorentz transformations, which is of central importance for
electromagnetism and special theory of relativity. $\ U(p,q)$ and
$\ SU(p,q)$ are defined similarly. For more details, we refer the
reader to [20,22]

\noindent We now define Homogeneous spaces and discuss some of their
properties:

\noindent\textbf{Definition 2.1.2:} \ We say that a Lie group $\ G$ is
\emph{represented as a Lie group of transformations of a
$\ C ^{\infty}$ manifold M (or has a left (Lie)- action on M)} if to
each  $\ g \in G $, there is associated a diffeomorphism from $\ M
$ to itself: \ $\ x \mapsto \psi_{g}(x) , x \in M $ such that $\
\psi_{gh} = \psi_{g} \psi_{h}$ for all $\ g, h \in G $ and $\
\psi_{e} = Id.$, Identity map of $\ M $, and if further-more $\
\psi_{g}(x) $ depends smoothly on the arguments $\ g $, $\ x $.
i.e. the map $\ (g, x) \mapsto \psi_{g}(x) $ is a smooth map from
$\ G \times M \rightarrow M $.

\noindent The Lie group $\ G$ is said to have a \emph{right action} on M if the
above definition is valid with the  property $\ \psi_{g} \psi_{h}
= \psi_{gh} $ replaced by $\ \psi_{g} \psi_{h} =  \psi_{hg} $.

\noindent If $\ G$ is any of the matrix Lie groups\\
$\ GL(n,R), O(n,R), O(p,q)$ or  \\ $\  GL(n, C), U(n), U(p,q)$ (where $\ p + q = n)$, then $\ G$
acts in the obvious way on the manifold $\ R^{n} $ or $\ R^{2n} = \ C^{n} $. In these cases, the elements of $\ G$ act as linear
transformations.

\noindent The action of a group $\ G$ is said to be \emph{transitive} if for
every two points $\ x, y $ of $\ M $, there exists an element $\ g
\in G $ such that $\ \psi_{g}(x) = y $.\\
\noindent\textbf{Definition 2.1.3:} \ A manifold on which a Lie group
acts transitively is called a \emph{homogeneous space} of the Lie
group.

\noindent In particular, any Lie group $\ G$ is a homogeneous space for itself
under the action of left multiplication. Here $\ G$ is called the
\emph{Principal left homogenous space} (of itself).  Similarly the
action $\ \psi_{g}(h) =  h g^{-1} $ makes $\ G$ into its own
\emph{Principal right homogeneous space.}

\noindent Let $\ x $ be any point of a homogeneous space of a Lie group $\ G$.
The \emph{isotropy group }(or stationary group) $\ H_{x} $ of the
point $\ x $ is the stabilizer of $\ x $ under the action of $\ G$ :
$\ H_{x} = \{ g \in G /  \psi_{g}(x) = x \} $. \\
We have the following lemma.

\noindent\textbf{Lemma 2.1.1:} All isotropy groups $\ H_{x}$  of
points $\ x $ of a homogeneous space are isomorphic.

\noindent\textbf{Proof :} Let $\ x $, y be any two points of the
homogeneous space. Let $\ g \in G $ be such that $\ \psi_{g}(x) =
y $. Then the map $\ H_{x} \rightarrow H_{y} $ defined by $\ h
\mapsto ghg^{-1} $  is an isomorphism. ( Here we have assumed the
left action).

\noindent We thus denote simply by H, the isotropy group of some (and hence
of every element modulo isomorphism)  element of $\ M $ on which $\ G$
acts on the left.\\
We now have the following theorem.

\noindent \textbf{Theorem 2.1.2:} There is a one- one
correspondence between the points of a homogeneous space $\ M $ of
the Lie group $\ G$, and the left cosets gH of H in $\ G$, where H is the
isotropy group and $\ G$ is assumed to act on the left.

\noindent\textbf{Proof: } Let $\ x_{0}$  be any point of the
manifold $\ M $. Then with each left coset $\ gH_{x_{0}}$  we
associate the point $\ \psi_{g}(x_{0})$  of $\ M $. Then this
correspondence is well- defined, i.e. independent of the choice of
representative of the coset, one - one and onto.

\noindent It can be shown under certain general conditions that the isotropy
group H is a closed sub group of $\ G$ , and the set $\ G/H $ with the
natural quotient topology can be given a unique (real) analytic
manifold structure such that $\ G$ is a Lie transformation group of $\
G/H $. Thus $\ M  \approx  G/H $.

\noindent\textbf{Examples of homogeneous spaces are:}

\noindent\textbf{1. Stiefel manifolds : } \ For each $\ n, k  ( k
\leq n)$, the Stiefel manifold $\ V_{n,k} $ has as its points all
orthonormal frames $\ x = (e_{1},e_{2}...,e_{k})$ of $\ k $
vectors in Euclidean $\ n $-space i.e. ordered sequences of $\ k $
orthonormal vectors in $\ R^{n} $. Then $\ V_{n,k} $ is embeddable
as a non- singular surface of dimension $\ nk - k (k+1)/2 $ in $\
R^{nk}$ and can be visualized as $\ SO(n) / SO(n-k) $. In
particular we have $\ V_{n,n} \cong O(n),  V_{n,n-1} \cong SO(n),
\  V_{n,1} \cong S^{n-1}$.

\noindent\textbf{ 2. Grassmannian manifolds :} \ The points of the
Grassmannian manifold $\ G_{n,k}$, are by definition, the $\ k $-
dimensional planes passing through the origin of $\ n
$-dimensional Euclidean space. This is a smooth manifold and it is
given by \\ $\ G_{n,k} \cong  O(n)/ O(k) \times O(n-k)$.
\\ We now define  symmetric spaces.

\noindent\textbf{Definition 2.1.4:} A simply connected manifold $\ M $
with a metric $\ g_{ab}$ defined on it, is called a
\emph{symmetric space} (\emph{symmetric manifold}) if for every
point $\ x $ of $\ M $, there exists an isometry (motion) $\ s_{x}
: M \rightarrow M $ with the properties that $\ x $ is an isolated
fixed point of it, and that the induced map on the tangent space
at $\ x $ reflects ( reverses ) every tangent vector at $\ x $
i.e. $\ \xi \mapsto  - \xi $. Such an isometry is called \emph{a
symmetry of M of the point $\ x $}.
\\ For every symmetric space, covariant derivative of Riemann
curvature tensor vanishes.
\\ For a homogeneous symmetric manifold $\ M $, let $\ G$ be the Lie
group of all isometries of $\ M $ and let H be the isotropy group
of $\ M $ with respect to left action of $\ G$ on $\ M $. Then , as we
have seen above, $\ M $ can be identified with $\ G/H $, the set of
left cosets of H in $\ G$. As examples of such spaces in general
relatively, we have the following space-times:

\noindent\textbf{Space of constant curvature with isotropy group
$\ H = SO(1,3)$}:

\noindent 1. Minkowski space $\ R^{4} $. \\
2. The de Sitter space \\
$\ S_{+} = SO(1,4) / SO(1,3)$. Here $\
S_{+}$ is homeomorphic to $\ R \times S^{3}$ and  the curvature
tensor $\ R $ is the identity operator on the space of bivectors $\
\Lambda^{2}(R^{4}) , R = Id $. \\
3. The anti- de Sitter space \\
$\ S = SO( 2,3) / SO(1,3)$. This
space is homeomorphic to $\ S^{1} \times R^{3}$ and  its universal
covering space is homeomorphic to $\ R^{4}$. Here curvature tensor
$\ R = - Id $.

\noindent Another example of symmetric space-time is the symmetric
space $\ M_{t} $ of plane waves. For these spaces the isotropy
group is abelian, and the isometry group   is soluble
(\emph{solvable}). (A group $\ G$ is called \emph{solvable} if it has
a finite chain of normal subgroups $\ \{e\} < G_{1} <...< G_{r }=
G $, beginning with the identity subgroup and ending with $\ G$, all
of whose factors $\ G_{i +1}/G_{i}$ are abelian). In terms of
suitable coordinates, the metric has the form \\ $\ ds^{2} =
2dx_{1} \ dx_{4}+ [ (cos \ t) x_{2}^{2} + (sin \ t)x_{3}^{2} ] \
dx_{4}^{2} + \ dx_{2}^{2} + \ dx_{3}^{2}$, $\ cos \ t \geq  \ sin
\ t $. The curvature  tensor is constant (refer [21]).

\noindent G\"{o}del universe [4] is also an example of
a homogeneous space but it is not a physically reasonable model
since it contains closed time like curve through every point.
We now turn our attention to Causal cones and cone preserving transformations.\\

\subsection { \emph{Causal cones and cone preserving transformations} }

We note that all genuine approaches to quantum gravity are free
of space-time metric while general relativity employs a Lorentzian
space-time metric. Hence, the question arises whether there exists
a structure which gets some features of light cones in a purely
topological manner. Motivated by the requirements on suitable
structures for a theory of quantum gravity,  new notions of causal
structure and cone structures were developed on a space-time $\ M $.
Here we describe these notions.\\
The definition of \emph{causal cone} is given as follows:\\
Let $\ M $ be a finite dimensional real Euclidean vector (linear)
space with inner product $\ < , > $.  Let $\ R^{+} $ be the set of
positive real numbers and $\ R^{+}_{0} = R^{+} \cup \{0\}$ . A
subset $\ C $ of $\ M $  is a \emph{cone} if $\ R^{+} C \subset C $ and
is a \emph{convex cone} if $\ C $, in addition, is a convex subset of
$\ M $. This means, if $\ x,y \in C $ and $\ \lambda \in [0,1] $,
then $\ \lambda x + (1-\lambda) y \in C $. In other words, $\ C $ is a
convex cone if and only if for all $\ x,y \in C $ and $\ \lambda,
\mu \in R^{+}, \lambda x + \mu y \in C.$ We call cone $\ C $ as
\emph{non- trivial }if $\ C \neq -C $. If $\ C $ is non-trivial, then
$\ C \neq \{0\} $ and $\ C \neq M $ .\\
We use the following notations:\\
i.  $\ M^{c} = C \cap - C $ \\
ii. $\ <C> = C - C = \{ x - y / \  x,y \in C \}$ \\
iii. $\ C^{*} = \{ x \in M / \forall \ y \in C, (x,y) \geq 0 \} $\\
Then $\ M^{c} $ and $\ <C> $ are vector spaces. They are called
the \emph{edge} and the \emph{span} of $\ C $. The set $\ C^{*}$  is a
closed convex cone called the \emph{dual cone} of $\ C $. This
definition  coincides with the usual definition of the dual space
$\ M^{*} $ of $\ M $ by using inner product ( , ). If $\ C $ is a
closed convex cone, we have $\ C^{**} = C $,  and $\ (C^{*} \cap -
C^{*} ) = < C > ^{\bot} ,$ where for $\ U \subset M , \ U^{\bot} =
\{ y \in M / \forall u \in U, (u, y ) = 0\} $.

\noindent\textbf{Definition 2.2.1:} Let $\ C $ be a convex cone in $\ M $.
Then $\ C $ is called \emph{generating} if $\ < C > = M $. $\ C $ is called
\emph{pointed} if there exists a $\ y \in M $ such that for all $\
x \in C - \{0 \} $, we have $\ (x,y) >  0 $. If $\ C $ is closed , it
is called \emph{proper} if $\ M^{c} = \{0\}$. $\ C $ is called
\emph{regular} if it is generating and proper. Finally, $\ C $ is
called \emph{self-dual}, if $\ C^{*} = C $.\\
If $\ M $ is an ordered linear space, the Clifford's theorem
[23] states that $\ M $ is directed if and only if $\ C $ is generating.\\
The set of interior points of $\ C $ is denoted by $\ C^{o} $ or $\ int
(C)$. The interior of $\ C $ in its linear span $\ <C> $ is called the
\emph{algebraic interior} of $\ C $ and is denoted by  \ alg int($\ C $).\\
Let $\ S \subset M$. Then the closed convex cone generated by $\ S $ is
denoted by \emph{Cone($\ S $)}:\\
Cone($\ S $) = \ closure of $\ \{ \displaystyle\sum_{finite} r_{s} s / s
\in S , r_{s} \geq 0 \} $.

\noindent If $\ C $ is a closed convex cone, then its interior $\ C^{o} $ is an
open convex cone. If $ \Omega $ is an open convex cone, then its
closure  $\ \overline{\Omega} = cl(\Omega)  $ is a closed convex
cone. For an open convex cone, we define the dual cone by \\
$\ \Omega^{*}  = \{ x \in v /  \forall  y \in \overline{\Omega} -
\{0\}  \ (x, y ) > 0 \} \ = int( \overline{\Omega^{*}})$ . \\
If $\ \overline{\Omega} $ is proper, we have $\ \Omega^{**} =
\Omega$ \\
We  now have the  following results: \ ( cf   [19,24])

\noindent\textbf{Proposition 2.2.1:} Let $\ C $ be a closed convex cone
in $\ M $. Then
the following statements are equivalent:\\
i.  $\ C^{o} $ is nonempty \\
ii. $\ C $ contains a basis of $\ M $.\\
iii. $\ < C > =  M $

\noindent\textbf{Proposition 2.2.2:} Let $\ C $ be a  nonempty closed
convex cone
in $\ M $. Then the following properties are equivalent :\\
i.  $\ C $ is pointed \\
ii. $\ C $ is proper\\
iii.    int ($\ C^{*}) \neq \phi $\\
As a  consequence, we have

\noindent\textbf{Corollary 2.2.3:} Let $\ C $ be a closed convex cone.
Then $\ C $ is proper if  and only if $\ C^{*}$ is generating.

\noindent\textbf{Corollary 2.2.4:} Let $\ C $ be a convex cone in $\ M $.
Then $\ C \in Cone (M) $ if and only if $\ C^{*} \in Cone (M) $.
Here $\ Cone(M)$ is the set of all closed regular convex cones in
$\ M $.

\noindent To proceed further along these lines, we need to make ourselves
familiar with more terminology and notations. The linear
automorphism group of a convex cone is defined as follows:\\
Aut ($\ C $) = $\{ a \in GL (M) / \alpha(C) = C \} $. GL ($\ M $) is
the group of invertible linear transformations of $\ M $. If $\ C $ is
open or closed, Aut ($\ C $) is closed in GL ($\ M $). In particular
Aut($\ C $) is a linear Lie group.

\noindent\textbf{Definition 2.2.2:} Let $\ G$ be a group acting linearly on
$\ M $. Then a cone $\ C \in M $ is called \emph{$\ G$- invariant} if
$\ G.C = C $. We denote the set of invariant regular cones in $\ M $ by
$\ Cone_{G}(M)$. A convex cone $\ C $ is called \emph{homogeneous} if
Aut ($\ C $) acts transitively on $\ C $.

\noindent For $\ C \in Cone_{G} (M)$, we have Aut ($\ C $) = Aut $\
(C^{o})$ and $\ C = \partial C \cup C^{o} = (C-C^{o}) \cup C^{o} $
is a decomposition of $\ C $ into Aut ($\ C $) - invariant subsets. In
particular a non-trivial closed \emph{regular} cone can never be
homogeneous. \noindent We have the following theorem:

\noindent\textbf{Theorem 2.2.5:} Let $\ G$ be a Lie group acting
linearly on the Euclidean vector space $\ M $ and $\ C \in
Cone_{G}(M)$. Then the stabilizer in $\ G$ of a point in $\ C^{o}$ is
compact.\\
\noindent \textbf{Proof:} Let $\ \Omega = C^{o}$, interior of a convex cone
$\ C $. Here, $\ C \in Cone_{G}(M)$, the set of G- invariant regular cones
in M. We first note that for every $\ v \in \Omega $, the set
$\ \textsl{U} = \Omega \bigcap (v -\Omega) $ is open (being intersection of two open sets),
non-empty (  $\frac{v}{2} \in \textsl{U} $) and bounded. Hence we can find closed balls $\ \overline{B_{r} ( \frac{v}{2}}) \subset \textsl{U} \subset \overline{B_{R} ( \frac{v}{2}} )$ ( by property of open sets in a metric space). Let $\ a \in Aut ( \Omega )^{v} = \{ b \in Aut(\Omega) / b.v = v \}$. Then $\ a.\Omega \subset \Omega $ and $\ a.v = v$. Thus we obtain $\ a ( \textsl{U} ) \subset \textsl{U}$. Hence, $\ a( \overline{B_{r} ( \frac{v}{2}} )) \subset a ( \textsl{U} ) \subset  \textsl{U} \subset \overline{B_{R} ( \frac{v}{2}} )$. Therefore, $\ a( \frac{v}{2}) = \frac{v}{2} $ implies $\ \parallel a \parallel \leq \frac{R}{r} .$ Thus $\ Aut ( \Omega )^{v} $ is closed and bounded, that is, compact.

\noindent In the abstract mathematical setting, cones  are described
using quasi-order relation  [25] as follows:

\noindent Let $\ M \neq 0 $ be a set and * be a mapping of $\ M
\times M $ into $\ \textsl{P}^{*}(M)$ (the set of all non-empty
subsets of $\ M $). The pair $\ ( M, * ) $ is called a
hypergroupoid. For $\ A, B \in \textsl{P}^{*}(M)$, we define $\ A*
B = \bigcup \{ a*b : a \in A, b \in B \}$.

\noindent A hypergroupoid $\ ( M,*)$ is called a hypergroup, if $\
(a*b)*c = a* (b*c) $ for all $\ a,b,c \in M$, and the reproduction
axiom, $\ a * M = M = M * a $, for any $ a \in M $,  is satisfied.

\noindent For a binary relation R on A and $\ a \in A $ denote $\ U_{R}(a) =
\{ b \in A/ <a , b> \in R \} $. A binary relation Q on a set A is
called quasiorder if it is reflexive and transitive. The set $\
U_{Q}(a)$  is called a cone of $\ a $. In the case when a
quasiorder Q is an equivalence,  $\ U_{Q}(A) = \{ x \in M /
\exists \ y \in A,  <x, y> \in \ Q \}$ for any $\ A \subseteq M $.
Analogously, for $\ B \subseteq A $ we set $\ U_{Q}(B) = \bigcap
\{ U_{Q}(a)/ a \in B \}$.

\noindent In the light of this definition, we shall observe in section 3 that causal cones
and K- causal cones fall in this category since causal relation $\ < $
and K-causal relation $\ \prec $ are reflexive and transitive.\\
In the literature, ( see for example [26,27,28]), cone
preserving mappings are defined as follows:

\noindent Let $\ \emph{\textbf{A}} = ( A, R ) $ and $\ \textbf{\emph{B}} = (B, S) $ be
quasi-ordered sets. A mapping $\ h: A \rightarrow B $ is called \emph{cone preserving}
if $\ h( U_{R}(a)) = U_{S}(h(a))$ for each $\ a \in A. $

\noindent To illustrate the concepts described above,  we consider the example
of the Minkowski space:

\subsection { \emph{Example  of a Forward Light cone in Minkowski space} }

\noindent \textbf{Note:} In the paper by Gheorghe and Mihul [29],
forward light cone is called \emph{`positive cone'}and is defined as follows:\\
Let M be a n-dimensional  real linear space. A causal relation of
M is a partial ordering relation $\ \geq $ of M with regard to
which M is \textit{directed }, i.e. for any $\ x, y \in M $ there
is $\ z \in M $ so that $\ z \geq x, z \geq y $. Then the positive
cone is defined as $\ C = \{x/ x \in M; x \geq 0 \}$

\noindent Let $\ p $ and $\ q $ be two positive integers and $\ n = p + q $. Let $\ M =
R^{n} $. We write elements of $\ M $ as $\ v = \left(
\begin{array}{r}
x \\
y
\end{array}
\right) $  with $\ x \in R^{p}$ and $\ y \in R^{q} $. For $\ p $ = 1 ,
$\ x $ is a real number.\\
We write projections $\ p_{r_{1}} $ and $\ p_{r_{2}}$  as $\
p_{r_{1}}(v)  = x $ and  $\ p_{r_{2}}(v)  = y $.

\noindent As discussed earlier, connected component of identity in $\ O(p,q)
$ denoted by  $\ O(p,q)_{o} = SO_{0}(p,q)= SO(p,q)_{0}$. Also Let
\\ $\ Q_{+r} = \{ x \in R^{n+1} / Q_{p +1, q} ( x , x )= r ^{2}\}
, r \in R ^{+} , p,q \in N , n = p + q \geq 1$. \\ Clearly, $\ O( p +
1, q)$ acts on $\ Q _{+r} $ . Let $\ \{ e_{1}, e_{2},
...e_{n}  \}$ be the standard basis for $\ R^{n} $. Then
we have the following result.

\noindent\textbf{Proposition 2.3.1: } For $\ p , q > 0 $, the group
$\ SO_{0}(p + 1, q)$ acts transitively on $\ Q _{+r} $. The
isotropy sub group at  $\ re_{1}$ is isomorphic to $\ SO_{0}
(p,q)$. As a
manifold,\\
$\  Q _{+r} \simeq  SO_{0} ( p + 1, q) / SO_{0} (p,q)$.\\
In particular for $\ n \geq 2, q = n-1 $ and $\ p = 1 $, we define
the semi algebraic cone $\ C $ in $\ R^{n}$ by \\ $\ C $ = $\ \{v \in R^{n
}/ Q_{1,q}(v, v) \geq 0, x\geq 0\}$ and set \\ $\ C^{*} = \Omega =
\{v \in R^{n} / Q_{1,q}(v, v)
> 0, x > 0 \}$. $\ C $ is called the \emph{forward light cone}
in $\ R^{n}$. We have
\noindent $\ M = \left(
\begin{array}{r}
x \\
y
\end{array}
\right) \in C $  if and only if $\   x \geq
 \parallel y \parallel $. \\
(Gheorghe and Mihul [29] state in Lemma 1 that \emph{There is
a norm $\parallel \parallel$ in $ \overline{M}$ ( a n-1
dimensional linear real space) so that: $\ Q = \{x/ x\in M;
\varepsilon x^{0} = \parallel \overline{x} \parallel\}, intC =
\{x/ x\in M;\varepsilon x^{0} > \parallel \overline{x}
\parallel\}$, where $\varepsilon = 1 $ if $\ (-1,\overline{0})$ is
not in $\ C $ and $\varepsilon = -1 $ if $\ (1,\overline{0})$ is not in
$\ C $)}.

\noindent Boundary of $\ C $ and $\ C^{o} $   are described  as
follows: $\ \partial C  =   \{ v \in R^{n} / \epsilon x =
\parallel y
\parallel \}, \ C^{o} = \{v \in R^{n} / \epsilon x > \parallel y \parallel \} $
where $\ \epsilon $ = 1 if $\ (-1, 0) $ is not in  $\ C $ and $\
\epsilon \neq 1 $ if $\  (1, 0) \in C $.
\\ If  $\ v \in C \cap -C $,  then $\ 0 \leq x \leq 0 $ and hence $\
x $ = 0. Then $\ \parallel y \parallel = 0 $ and thus $\ y = 0$.
Thus $\ v = 0$  and $\ C $ is proper. \\ For $\ v, v' \in C $, we
calculate
\\ $\ (v, v') = (v',v) = x^{'}x + (y^{'} , y) \geq \parallel y'
\parallel
\parallel  y \parallel + (y',y) \geq 0$. Thus $\ C \subset C^{*}$.

\noindent Conversely, let $\  v = \left(
\begin{array}{r}
x \\
y
\end{array}
\right) \in C^{*}$. Then testing against $\ e_{1}$, we get $\ x
\geq 0$. We may assume $\ y \neq 0 $. Define $\ \omega $  by $\
p_{r_{1}} (\omega) = \parallel y \parallel $ and $\ p_{r_{2}}
(\omega)  = -y$. Then $\ \omega \in C $ and  $\ 0 \leq (w, v) = x
\parallel y \parallel - \parallel y
\parallel^{2} = (x - \parallel y \parallel) \parallel y \parallel
$. Hence $\ x \geq \parallel y \parallel$. Therefore $\ y \in C $
and thus $\ C^{*} \subset C$. So $\ C = C^{* }$ and $\ C $ is
self-dual. Similarly, we can show that $\ \Omega $ is self dual.

\noindent Moreover, the forward light cone $\ C $ is invariant under the usual
operation of $\ S O_{o}(1,q)$ and under all dilations, $\ \lambda
I_{n},  \lambda > 0$. ($\ I_{n}$  is the $\ n \times \ n $
identity matrix). We now prove that the group $\ S O_{o}(1,q)
R^{+} I_{q+1}$ acts transitively on $\ \Omega = C^{o} $ if $\ q
\geq 2 $ ( $\ q = 3$ for Minkowski space). Thus $\ \Omega $ will be
homogeneous.For this we prove that $\ \Omega  = SO_{o} (1,q) R^{+} \left(
\begin{array}{r}
1 \\
0
\end{array}
\right)$. \\
Using  \\
$\ a_{t} = \left(
\begin{array}{rrr}
cosh(t) & sinh(t) & 0 \\
sinh(t) & cosh(t) & 0 \\
0 & 0 & I_{n-2}
\end{array}
\right) \in SO_{o} (1,q)$,  we get \\
$\ a_{t}\left(
\begin{array}{r}
\lambda \\
0
\end{array}
\right) = \lambda^{t} (cosh(t), sinh(t), 0, \cdots ,0) $ for all
$\ t \in R$. Let $\ S^{q-1}$ denote a unit sphere in $\ R^{q}$.
Now $\ SO(q)$ acts transitively on $\ S^{q-1} $ and $\ \left(
\begin{array}{rr}
1 & 0 \\
0 & A
\end{array}
\right) \in S O_{o}(1,q)$ \\ for all $\ A \in SO(q)$. Hence the
result follows by noting the fact that coth(t) runs through $\
(1,\infty)$ as t varies in $\ (0,\infty)$.

\section{Causal Structure of Space-times, Causality Conditions and Causal group}

\subsection{ \emph{Causal Structure and K- Causality} }

In this section, we begin with basic definitions and properties of
causal structure of space-time. Then we define different causality
conditions and their hierarchy. Furthermore we discuss causal
group and causal topology on space-time in general, and treat
Minkowski space as a special case. We take a space-time ($\ M $,
g) as a connected $\ C^{2} $ - Hausdorff  four dimensional
differentiable manifold which is paracompact and admits a
Lorentzian  metric  g of signature (-, +, +, + ). Moreover, we
assume that the space-time is space and time oriented.

\noindent We say that an event $\ x $ \emph{chronologically} precedes
another event $\ y $, denoted by $\ x \ \ll \  y $ if there is a
smooth future directed  timelike curve from $\ x $ to $\ y $ . If
such a curve is non-spacelike, i.e., timelike or null , we say
that $\ x $ causally precedes $\ y $  or $\ x < \  y $. The
chronological future $\ I^{+}(x) $ of $\ x$ is the set of all
points $\ y$ such that $\ x \ \ll \ y $ . The chronological past
$\ I^{-}(x) $ of $\ x$ is defined dually. Thus

     $\ I^{+}(x)  = \{  y \in \ M /  x \ \ll \  y \} $ \ and

     $\ I^{-}(x)  = \{  y \in \ M /  y \ \ll \  x \} $.

\noindent The causal future and causal past for $\ x $  are defined
similarly :

    $\ J^{+}(x)  = \{  y \in \ M /  x \ < \  y \} $ \ and

      $\ J^{-}(x)  = \{  y \in \ M /  y \ < \  x \} $

\noindent As Penrose [30] has proved, the relations $\ \ll $  and $\ < $ are
transitive. Moreover,\\ $\  x \ll \ y $ and $\ y < z $ or $\ x < y
$ and $\ y \ll z $ implies $\ x \ll z $. Thus  $\ \overline
{I^{+}(x)} = \ \overline {J^{+}(x)} $   and also $\  \partial
I^{+}(x) = \ \partial J^{+}(x) $, where for a set $\ X \subset \ M
$, $\ \overline{X} $ denotes closure of $\ X $ and  $\ \partial X
$ denotes topological boundary of $\ X $. The chronological future
and causal future of any set $\ X \subset \  M $ is defined as

    $\ I^{+}(X) = \displaystyle\bigcup_{ x \in  X } I^{+}(x) $ and

    $\   J^{+}(X) = \displaystyle\bigcup_{ x \in  X } J^{+}(x) $ \\
The chronological and causal pasts for subsets of $\ M $ are
defined similarly.

\noindent An ordering which is reflexive and transitive is called
quasi - ordering. This ordering was developed in a generalized
sense by Sorkin and Woolgar [8] and these concepts were
further developed by Garcia Parrado and Senovilla [31,32]
and S. Janardhan and Saraykar [9] to prove many interesting
results in causal structure theory in General Relativity.

\noindent In the recent paper, Zapata and Kreinovich [28]  call
chronological order as open order and causal order as closed order
and prove that under reasonable assumptions, one can uniquely
reconstruct an open order if one knows the corresponding closed
order. For special theory of relativity, this part is true and
hence every one-one transformation preserving a closed order
preserves open order and topology. This fact in turn implies that
every order preserving transformation is linear. The conserve part
is well known namely, the open relation uniquely determines both
the topology and the closed order.

\noindent We now introduce the concept of K-causality and give causal properties
of space-times in the light of this concept. For more details we refer the reader to
[9], [11,12]   and [31,32].

\noindent\textbf{Definition 3.1.1:} \emph{$\ K^{+}$} is the smallest
relation containing $\ I^{+}$ that is topologically closed  and
transitive. If  $\ q $ is in $\ K^{+}(p)  $ then we write $\ p \prec q
$.

\noindent That is, we define the relation $\ K^{+}$, regarded as a subset of
$\ M \times M $, to be the intersection of all closed subsets $\ R
\supseteq I^{+} $ with the property that $\ (p, q)  \in  \ R $ and
$\ (q, r) \in \ R $   implies $\ (p, r) \in \ R $. ( Such sets R
exist because $\ M \times M $ is one of them.) One can also
describe $\ K^{+}$ as the closed-transitive relation generated by
$\ I^{+}$.

\noindent\textbf{Definition 3.1.2:}\ An open set O is \emph{K-causal}
iff the relation `$\ \prec $' \ induces a reflexive partial
ordering on O. i.e. $\ p \prec q $ and $\ q \prec p $ together
imply $\ p = q $.

\noindent If we regard $\ C^{o} $ as the interior of future light
cone in a Minkowski space-time ($\ p = 1, q = 3$ ), then under
standard chronological structure $\ I^{+} , M (a,b)$ becomes $\
I^{-}(b) \cap I^{+}(a)$. As it is well known, such sets form a
base for Alexandrov topology and since Minkowski space-time is
globally hyperbolic and hence strongly causal, Alexandrov topology
coincides with the manifold topology (Euclidean topology). Thus,
lemma 2 of [9] is a familiar result in the language of
Causal structure theory.

\noindent Analogous to usual causal structure, we  defined in [9] strongly
causal and future distinguishing space-times with respect to $\
K^{+}$ relation.

\noindent\textbf{Definition 3.1.3:} \  A $\ C^{0} $ - space-time $\ M $
is said to be \emph{strongly causal at $\ p $ with respect to $\
K^{+}$}, if $\ p $ has arbitrarily small K - convex open
neighbourhoods. \\
Analogous definition would follow for $\ K^{-} $.\\
$\ M $ is said to be \emph{strongly causal with respect to $\
K^{+}$}, if it is strongly causal with respect to $\ K^{+}$ at
each and every point of it. Thus, lemma 16 of [8] implies that
K-causality implies strong causality with respect to $\ K^{+} $.

\noindent\textbf{Definition 3.1.4:} \ A $\ C^{0} $-  space-time $\ M $
is said to be \emph{K-future distinguishing} if for every $\ p
\neq q , K^{+}(p) \neq K^{+}(q) $. \ \emph{K-past distinguishing}
spaces can be defined analogously.

\noindent\textbf{Definition 3.1.5:} \ A $\ C^{0} $- space-time $\ M $
is said to be \emph{K-distinguishing }if it is both K-future and
K-past distinguishing.

\noindent Analogous result would follow for $\ K^{-} $. Hence, in  a $\
C^{0}$ - space-time  $\ M $, strong causality with respect to K
implies K-distinguishing.

\noindent\textbf{Remark :} \ K-conformal maps preserve K-
distinguishing,  strongly causal with respect to $\ K^{+}$ and
globally hyperbolic properties.

\noindent\textbf{Definition 3.1.6:} \ A $\ C^{0} $- space-time $\ M $
is said to be \emph{K-reflecting} if  \\ $\ K^{+}(p) \supseteq
K^{+}(q) \Leftrightarrow \  K^{-}(q) \supseteq K^{-}(p) $.\\
However, since the condition $\ K^{+}(p) \supseteq  K^{+}(q) $
always implies $\ K^{-}(q) \supseteq K^{-}(p) $ because of
transitivity and $\ x \in K^{+}(x) $, and vice versa, a $\ C^{0}$
- space-time  with K-causal condition is always K-reflecting.
Moreover, in general, K-reflecting need not imply reflecting.
Since, any K-causal space-time is K-reflecting, any non-reflecting
open subset of the space-time will be K-causal but non-reflecting.

\noindent We now give the interesting hierarchy of K-causality
conditions as follows: \\ We have proved that strong causality
with respect to $\ K^{+} $ implies K-future distinguishing. Thus,
K-causality $\ \Rightarrow $ \ strongly causality with respect to
K  $\ \Rightarrow $ \ K - distinguishing.

\noindent Since  a K- causal space-time is always K-reflecting, it follows
that the K-causal space-time is K-reflecting as well as
K-distinguishing. In the classical causal theory, such a
space-time is called causally continuous [33]. (Such space-times
have been useful in the study of topology change in quantum
gravity [34]). Thus if we define K-causally continuous space-time
analogously then we get the result that a K-causal $\ C^{0}$ -
space-time   is K-causally continuous. Moreover, since $\
K^{\pm}(x) $ are topologically closed by definition, analogue of
causal simplicity is redundant and causal continuity (which is
implied by causal simplicity) follows from K-causality.

\noindent In [9],we proved the following theorems.
Here we recall their proofs for the sake of completeness.

\noindent\textbf{Theorem 3.1.1 :} \ Let V be a globally hyperbolic $\
C^{0}$- space-time. If $\ S \subseteq V $ is compact then $\
K^{+}(S) $ is closed.

\noindent\textbf{Proof :} \ Let $\ S \subseteq V $ be compact. Let
$\ q \in \ cl(K^{+}(S))$. Then there exists a sequence $\ q_{n}$
in $\ K^{+}(S) $ such that $\ q_{n}$ converges to q. Hence there
exists a sequence $\ p_{n}$ in S  corresponding to $\ q_{n}$ and
future directed K- causal curves $\ \Gamma_{n}$ from $\ p_{n}$ to
$\ q_{n}$. Then $\ p_{n}$ has a subsequence $\ p_{n_{k}}$
converging to $\ p\in S $ since S is compact, which gives a
subsequence $\ \Gamma_{n_{k}}$ of future directed K-causal curves
from $\ p_{n_{k}}$ to $\ q_{n_{k}}$ where $\ p_{n_{k}}$ converges
to p and $\ q_{n_{k}}$ converges to q. Define P = $\  \{p_{n_{k}} , p\} $
and Q = $\  \{q_{n_{k}} , q \}$. Then P and Q are compact subsets of V. Hence
the set \emph{\textbf{C}} of all future
directed K-causal curves from P to Q is compact. Now, $\ \{ \Gamma_{n_{k}}\}  $
is a subset of \emph{\textbf{C}}. Thus, $\ \{ \Gamma_{n_{k}} \}
$ is a sequence in a compact set and hence has a convergent
subsequence say $\ \Gamma_{n_{k_{l}}} $ of future directed
K-causal curves from $\ p_{n_{k_{l}}} $ to $\ q_{n_{k_{l}}} $
where $\ p_{n_{k_{l}}} $ converges to p and  $\ q_{n_{k_{l}}} $
converges to q .Let $\ \Gamma $ be the Vietoris limit of $\
\Gamma_{n_{k_{l}}} $. Then $\ \Gamma $ is a
future directed K-causal curve from p to q . Since $\ p \in S $,
we have, $\ q\in K^{+}(S)$. Hence $\ cl(K^{+}(S)) \subseteq
K^{+}(S)$. Thus $\ K^{+}(S) $ is closed.

\noindent The next two theorems show that in a globally hyperbolic $\ C^{0}
$ - space-time V , it is possible to express $\ K^{+}(x) $ in
terms of $\ I^{+}(x) $.

\noindent\textbf{Theorem 3.1.2 :} \ If V is a globally hyperbolic $\
C^{0} $ - space-time, then $\ K^{+}(p) = cl (int (K^{+}(p)),  p \in V $.

\noindent\textbf{Proof :} \ Let V be globally hyperbolic. It is
enough to prove that  $\ K^{+}(p) \subseteq \ cl ( int (K^{+}(p)),
 p \in \ V $. For this we show that $\ cl ( int (K^{+}(p)) $
is closed with respect to transitivity. So, let $\ x, y, z  \in \
cl ( int (K^{+}(p)) $ such that $\ x \prec \ y $ and $\ y \prec \
z $. We show that $\ x \prec \ z $. Since $\ x, y, z $ are limit
points of $\ int ( K^{+} (p))$, there are sequences $\ \{x_{n}\},
\  \{y_{n}\} , \ \{z_{n}\} $ in $\ int ( K^{+} (p) ) $ such that
$\ x_{n} \rightarrow \ x, \  y_{n} \rightarrow \ y, \  z_{n}
\rightarrow \ z $. Using first countability axiom , we may assume,
without loss of generality, that these sequences are linearly
ordered in the past directed sense [13]. Thus, for sufficiently large n, we
can assume that $\ x_{n} \prec \ y_{n}$ and $\ y_{n} \prec \ z_{n}
$. Since $\ x_{n}, \ y_{n}, \ z_{n} \in \ K^{+}(p) $, by
transitivity, $\ x_{n} \prec \ z_{n}$ for sufficiently large n. We
claim that $\ x \prec \ z $. Let $\ x $ be not in $\ K^{-}(z) $.
Then as $\ K^{-}(z) $ is closed,using local compactness, there
exists a compact neighbourhood N of x such that $\ N \cap K^{-}(z)
= \emptyset $, and so, $\ z $ is not in $\ K^{+}(N) $. Now
as V is globally hyperbolic and N is compact, $\ K^{+}(N) $ is closed. Hence,  there exists a K-convex
neighbourhood $\ N^{'} $ of z such that $\ N^{'} \cap K^{+}(N) =
\emptyset $, which is a contradiction as $\ x_{n} \prec z_{n}$ for
large n. Hence,  $\ x \prec \ z $. Thus, $\ cl ( int (K^{+}(p))$
is closed with respect to transitivity. Since, by definition, $\
K^{+}(p) $ is the smallest closed set which is transitive, we get,
$\ K^{+}(p) \subseteq \ cl ( int (K^{+}(p)) $. Hence  $\ K^{+}(p)
= \ cl ( int (K^{+}(p)) $. Similarly, $\ K^{-}(p) = \ cl ( int
(K^{-}(p)) $.

\noindent\textbf{Theorem 3.1.3 :} \ If V is a globally hyperbolic $\
C^{0} $ - space-time then \\
 $\ int (K^{\pm}(x)) = I^{\pm}(x) , \ x \in V $.

\noindent\textbf{Proof :} \ Let V be globally hyperbolic and $\ x
\in \ V $. That $\ I^{+}(x) \subseteq \ int (K^{+}(x)) $ is
obvious by definition of $\ K ^{+}(x) $. To prove the reverse
inclusion, we prove that, if $\ x \prec \ y $ then there exists a
K-causal curve from $\ x $ to $\ y $ and if $\ y \in \ int(
K^{+}(x) )$, then this curve must be a future-directed time-like
curve.

\noindent Let $\ x \prec \ y $ and there is no K-causal curve from $\ x $ to
$\ y $. Then image of [0,1] will not be connected, compact or
linearly ordered. This is possible, only when a point or a set of
points has been removed from the compact set $\ K^{+}(x) \cap
K^{-}(y) $ , that is, when some of the limit points have been
removed from this set , which will imply that this set is not
closed.

\noindent But, since V is globally hyperbolic, $\ K^{+}(x) \cap K^{-}(y) $
is compact and hence closed. Hence, there must exist a K-causal
curve from $\ x $ to $\ y $.

\noindent Suppose, $\ y \in \ int (K^{+}(x)) $. Then, there exists a
neighbourhood $\ I^{+}(p) \cap I^{-}(q) $ of $\ y $ such that  $\ y \in
\ I^{+}(p) \cap I^{-}(q) \subseteq K^{+}(x) $. To show that a
K-causal curve from $\ x $ to $\ y $ is time-like, it is enough to
prove that $\ x $ and $\ y $ are not null-related, that is, there
exists a non-empty open set in $\ K^{+}(x) \cap \ K^{-}(y)$.

\noindent Consider, $\ I^{+}(p) \cap I^{-}(q) \cap I^{+}(x) \cap I^{-}(y) $,
which is open. Take any point say $\ z $, on the future-directed
time-like curve from $\ p $ to $\ y $ . \\ Then, $\ z \in I^{+}(p) \cap
I^{-}(q) \cap I^{+}(x) \cap I^{-}(y)  \subseteq \ K^{+}(x) \cap \
K^{-}(y) $. ( Here, $\ z \in \ I ^{+}(x) $  because, if $\ x $ and $\ z $
are null-related then $\ K^{+}(x) \cap \ K^{-}(z) $ will not
contain an open set. But $\ I^{+}(p) \cap \ I^{-}(z) \subseteq \
K^{+}(x) \cap \ K^{-}(z) $ ).  That is, $\ K^{+}(x) \cap \
K^{-}(y) $ has a non-empty open subset. Hence, $\ x $  and  $\ y $ are not
null-related, and so, the K-causal curve from $\ x $  to $\  y $ is
time-like. That is, $\ y\in \ I^{+}(x) $. Thus, $\ int (K^{+}(x))
\subseteq I^{+}(x) $ which proves that $\ int (K^{+}(x)) = \
I^{+}(x) $. Similarly, we can prove that $\ int (K^{-}(x)) = \
I^{-}(x) $.

\noindent We now discuss important contribution by Minguzzi [12].
\noindent We recall that, $\ (M, g)$ is stably causal if there is $\ g^{'} >
g $ with $\ (M, g^{'})$ causal. Here $\ g^{'} > g $ if the light
cones of $\ g^{'} $ are everywhere strictly larger than those of
$\ g $. Equivalence of K-causality and stable causality uses the
concept of  \emph{ compact stable causality}  introduced in [11]. A
space-time is compactly stably causal if for every compact set,
the light cones can be widened on the compact set while preserving
causality. In [12], Minguzzi proved that K-causality implies
compact stable causality, and he also gave examples which showed
that the two properties differ. It will not be out of place here
to mention relationship between stable causality, Seifert future
$\ J^{+}_{s}(x)$  , almost future $\ A^{+}(x)$ and smooth and
temporal time functions.

\noindent For detailed discussion of these concepts, we refer the reader
to [6] and a more recent review by M. Sanchez [35].\\
\noindent Seifert future is defined as
$\ J^{+}_{s}(x) = \bigcap_{g^{'}> g} J^{+} (x, g^{'})$.

\noindent Then, $\ J^{+}_{s}$ is closed, transitive and contains $\ J^{+}$.
The space-time is stably causal if and only if $\ J^{+}_{s}$ is
anti-symmetric and hence a partial ordering on $\ M. $  (for
proof, we refer to Seifert [36]).

\noindent Another causality condition related to Seifert future is Almost future [37],
which is defined as follows : \\
An event $\ x $ almost causally precedes another event $\ y $, denoted by $\ x A y $, if
for all $\ z \in  I^{-}(x) , I^{+}(z) \subseteq I^{+}(y)$. We now
define $\ A^{+}(x) = \{ y \in M / x A y \} $. $\ A^{-}(x)$ is
defined similarly. It is clear that $\ y \in  A^{+}(x)$ if and
only if $\ x \in A^{-}(y)$. A space-time is called W-causal if $\
x \in A^{+}(y)$ and $\ y \in A^{+}(x) $ implies $\ x = y $ for all
$\ x ,y \in M $.

\noindent It is proved in [6] [Prop.4.12] that the almost future $\
A^{+}(x)$ is closed in the manifold topology for all $\ x \in M $.
Moreover [Prop.4.15], for all $\ x \in M,  A^{+}(x) \subseteq  J^{+}_{S} $.
In general, stable causality implies W-causality, though converse is not always true. Also, there is an
interesting relationship between stable causality and existence of
time functions.

\noindent We give the following definition : Let $\ (M, g)$ be a
space-time. A (non-necessarily continuous) function $\ t: M
\rightarrow R $ is:\\
(i) A generalized time function if t is strictly increasing on any
future-directed causal curve $ \gamma $. \\
(ii) A time function if t is a continuous generalized time
function. \\
(iii) A temporal function if t is a smooth function with
past-directed time-like gradient $ \nabla t $.

\noindent Then, we have the following theorem : \\
\noindent \textbf{Theorem 3.1.4:}   For a space-time $\ (M, g)$ the
following properties are equivalent: \\
(i) To be stably causal.\\
(ii) To admit a time function t \\
(iii) To admit a temporal function T

\noindent See [35] for the proof  and  more detailed discussion on
Causal hierarchy. See also Joshi [6], section 4.6, for a general
discussion on causal functions and relationship with stably causal
space-times. Coming back to relation $\ K^{+}$, we recall that $\
K^{+}$ is the smallest closed and transitive relation which
contains $\ J^{+}$. A space-time is K-causal if $\ K^{+}$ is
anti-symmetric. By definition, $\ K^{+}$ is contained in $\ J^{+}
_{S} $ , but they do not coincide. However, K-causality is
equivalent to stable causality and in this case $\ K^{+} =
J^{+}_{S} $. In [12], Minguzzi proves the equivalence of
K-causality and stable causality. For this, he develops a good
deal of new terminology and proves a series of lemmas, and uses
results proved in earlier papers [11,38,39]. Once this equivalence is
proved, it also follows that in a K-causal space-time, $\ K^{+}$
relation coincides with the Seifert relation, as mentioned above .

\noindent This equivalence, which follows after a laborious work extended
over a series of four papers, considerably simplifies the
hierarchy of Causality conditions, which now reads as :\\
\noindent Global hyperbolicity $\ \Rightarrow $ Stably causal $\
\Leftrightarrow $ K-causality  $\ \Rightarrow $ Strong causality
\\
$\ \Rightarrow $ K - Distinguishing.

\subsection{ \emph{Causal Groups and Causal Topology}}

We now discuss causal groups and causal topology and
then compare these notions with those in section 2.\\
If $\ R^{n}$ is a directed set with respect to a certain partial
ordering relation `$\ \geq $' of $\ R^{n}$, then such a relation
is called a \emph{Causal relation}. Thus in a globally hyperbolic
space-time (or in a Minkowski space time) $\ J^{+}$ and $\ K^{+}$
are causal relations (In a $\ C^{2}$ globally hyperbolic
space-time, $\ J^{+} = K^{+}$, whereas in a $\ C^{0}$ - globally
hyperbolic space-time, only $\ K^{+}$ is valid). The \emph{Causal
group} $\ G$ relative to causal relation is then defined as the group
of permutations $\ f : R^{n} \rightarrow R^{n}$ which leaves the
relation `$\ \geq $' invariant. i.e. $\ f(x) \geq f(y) $ if and
only if $\ x \geq y $. Such maps are called causal maps. They
preserve causal order. These maps are special cases of  cone
preserving maps defined in section 2.
We define the K-causal map and discuss their properties briefly.\\
A K- causal map is a causal relation which is a homeomorphism
between the two topological spaces and at the same time preserves
the order with respect to $\ K^{+}$. To define it, we first define
an order preserving map with respect to $\ K^{+} $:

\noindent\textbf{Definition 3.2.1:} Let V and W be $\ C^{0}$- space
- times. A mapping $\ f: V \rightarrow W $ is said to be
\emph{order preserving with respect to} $\ K^{+} $ or simply
\emph{order preserving }if whenever $\ p,q \in V $ with $\ q \in
K^{+}(p),$ we have $\ f(q) \in K^{+}(f(p))$. i.e., $\ p \prec q $
implies $\ f(p) \prec \ f(q) $.

\noindent\textbf{Definition 3.2.2:}\ Let V and W be $\ C^{0}$- space -
times. A homeomorphism $\ f: V \rightarrow W $ is said to be
\emph{K-causal} if $\ f $ is order preserving.

\noindent\textbf{Remark :}\ In general K-causal maps and causal
maps defined by A. Garcia-Parrado and J.M. Senovilla [31,32] are
not comparable as $\ r \in K^{+}(p) $ need not imply that $\ r \in
I^{+}(p) $ (refer figure 1 of [9]).

\noindent Using the definition of K-causal map, we now prove a series of
properties which follow directly from the definition. We give
their proofs for the sake of completeness:

\noindent \textbf{Proposition 3.2.1 :} \ A homeomorphism $\ f: V \rightarrow W $  is order
preserving iff
\\ $\ f((K^{+}(x))\subseteq K^{+}(f(x)), \forall \ x \in V $.

\noindent \textbf{Proof :} Let $\ f: V \rightarrow W $ be an order preserving homeomorphism
and let $\ x \in V $. Let $\ y \in f(K^{+}(x))$. Then $\ y = f(p),
\  x \prec p $ which implies $\ f(x) \prec \ f(p)$ as $\ f $is
order preserving. i.e., $\
f(x)\prec y $ or  $\ y \in \ K^{+}(f(x))$. \\
Hence $\ f(K^{+}(x)) \subseteq \ K^{+}(f(x)), \ \forall \ x \ \in \ V $.\\
Conversely let $\ f: V \rightarrow W $ be a homeomorphism such
that \\ $\ f((K^{+}(x)) \subseteq  \ K^{+}(f(x)), \ x \in \  V $. \\
Let $\ p \prec q $. Then  $\  f(q) \in \ f(K^{+}(p))$. By
hypothesis, this gives  $\ f(q) \in \ K^{+}(f(p))$. Hence $\ f(p)
\prec \ f(q)$. Thus if $\ f $ is a K-causal map then $\
f(K^{+}(x)) \subseteq \ K^{+}(f(x)),  \forall \ x \in V $.

\noindent Similarly we have the property:

\noindent \textbf{Proposition 3.2.2 :} \ If $\ f: V \rightarrow
W $ be a homeomorphism then $\ f^{-1}$ is order preserving iff $\
K^{+}(f(x))\subseteq f((K^{+}(x)), \ x \in V $.

\noindent \textbf{Proof:} We now define, for $\ S\subseteq V,  $   $\  K^{+}(S)=  \bigcup_{x
\in S } K^{+}(x)  $.

\noindent In general, $\ K^{+}(S) $ is neither open nor closed. We
shall show that in a globally hyperbolic $\ C^{0} $  space-time, if S is compact, then $\ K^{+}(S) $ is closed.
However at present, we can prove the following property:

\noindent \textbf{Proposition 3.2.3 :} \ If $\ f: V \rightarrow W $  is an order
preserving homeomorphism then $\ f((K^{+}(S))\subseteq
K^{+}(f(S)), \ S\subseteq  V $.

\noindent \textbf{Proof:} If $\ f: V \rightarrow W $ be an order preserving
homeomorphism and \\ $\ S\subseteq V $ then by definition, $\
K^{+}(S)= \bigcup_{x\in S} K^{+}(x)$. Let $\ y \in f(K^{+}(S)) $.
Then there exists $\ x $ in S such that $\ y
 \in f(K^{+}(x))$. This gives $\ y \in K^{+}(f(x))$.\\
 i.e., $\ y \in K^{+}(f(S))
$. Hence $\ f(K^{+}(S)) \subseteq  K^{+}(f(S))$.

\noindent Analogously we have,

\noindent \textbf{Proposition 3.2.4 :} \ If $\ f: V \rightarrow W $ be a
homeomorphism, and $\ f^{-1}$ is order preserving then $\
K^{+}(f(S)) \subseteq \ f(K^{+}(S)), \ S \subseteq \ V $.

\noindent We know that causal structure of space-times is given by its
conformal structure. Thus, two space-times have identical
causality properties if they are related by a conformal
diffeomorphism. Analogously, we expect that a K- conformal map
should preserve K- causal properties. Thus we define a K-
conformal map as follows.

\noindent\textbf{Definition 3.2.3:}\ A homeomorphism $\ f: V
\longrightarrow W $ is said be K-conformal if both $\ f $ and $\
f^{-1} $ are K-causal maps.

\noindent\textbf{Remark :}\ A K-conformal map is a causal
automorphism in the sense of E.C.Zeeman [40].

\noindent This definition is similar to \emph{chronal / causal
isomorphism} of Zeeman [40], Joshi [6] and Garcia - Parrado and
Senovilla [31,32].

\noindent Combining the above properties , we have the following:\\
\noindent\textbf{Proposition 3.2.5 :} \ If $\ f: V \rightarrow \ W $ is K-conformal then $\
f(K^{+}(x)) = K^{+}(f(x)), \forall \ x \in V $.

\noindent By definition, K- conformal map will preserve different K-
causality conditions. If a map is only
K- causal and not K- conformal, then we have the following
properties:

\noindent\textbf{Proposition 3.2.6 :} \ If  $\ f: V \rightarrow W $ is a
K-causal mapping and W is K-causal, so is V.

\noindent\textbf{Proof :}  Let $\ f: V \rightarrow W $ be a K-causal map and W be
K-causal. Let $\ p \prec q $ and  $\  q \prec p, \ p, \ q \in V $.
Then $\ f(p), \ f(q) \ \in W $ such that $\ f(p) \prec f(q)$ and
$\ f(q) \prec f(p)$ as $\ f $ is order preserving. Therefore $\
f(p) = f(q) $ since W is K-causal. Hence $ \ p = q $.

\noindent Analogous result would follow for $\ f^{-1} $.

\noindent In addition, a K-causal mapping takes K-causal curves to K-causal curves. This is
given by the property:\\
\noindent\textbf{Proposition 3.2.7 :} \ If V be a K-causal space-time and $\ f:V \rightarrow W $ be a K-causal mapping, then $\ f $
maps every K-causal curve in V to a K-causal curve in W.

\noindent\textbf{Proof :}Let $\ f: V \rightarrow W $  be a K-causal map. Therefore $\ f $
is an order preserving homeomorphism. Let $\ \Gamma $ be a
K-causal curve in V. Then  $\ \Gamma $ is
connected, compact and linearly ordered. Since $\ f $ is
continuous, it maps a connected set to a connected set and a
compact set to a compact set. Since $\ f $ is order preserving and
$\ \Gamma $ is linearly ordered, $\ f(\Gamma) $ is a K-causal
curve in W. Analogous result would follow for $\ f^{-1} $.

\noindent From the above result we can deduce the following:\\
\noindent\textbf{Proposition 3.2.8 :} \ If $\ f$ be a
K-causal map from V to W, then for every future directed K-causal
curve $\ \Gamma $ in V, any two points $\ x, \ y \in \ f(\Gamma) $
satisfy $\ x \prec y \ or \ y \prec x $.

\noindent\textbf{Definition 3.2.4:}\ \ Let V and W be two $\ C^{0}$ -
space-times. Then W is said to be \emph{K-causally related }to V
if there exists a K-causal mapping $\ f $ from V to W. i.e., $\ V
\prec_{f}W $.

\noindent The following property follows easily from this
definition, which shows that the relation $\ `\prec_{f}'$ is
transitive also.

\noindent \textbf{Proposition 3.2.9 :} \ If $\ V \prec_{f}W $ and $\ W \prec_{g}U $ then $\ V \prec_{g \circ f}U $.

\noindent \textbf{Proposition 3.2.10 :} \ If $\ f: V \longrightarrow W $ is a K-causal map then
$\ C \subseteq V $ is K-convex if $\ f(C)$ is a K-convex subset of
W.

\noindent \textbf{Proof :} Let $\ f: V \longrightarrow W $ be K-causal and $\ f(C) $ be a
K-convex subset of W. Let $\ p , q \in C $ and $\ r \in V $ such
that $\ p \prec r \prec q $. Since $\ f $ is order preserving we
get  $\ f(p) \prec f(r) \prec f(q) $ where $\ f(p),f(q) \in f(C) $
and $\ f(r) \in W $. Since $\ f(C) $ is K-convex, $\ f(r) \in f(C)
$. i.e., $\ r \in C $. Hence C is a K-convex subset of V.

\noindent\textbf{Remark :}\ Concept of a convex set is needed to
define \emph{strong causality}, as we shall see below.

\noindent We now discuss briefly the algebraic structure of the set of all
K-causal maps from V to V. We define the following:

\noindent\textbf{Definition 3.2.5:} \ If V is a $\ C^{0}$ - space -
time then \emph{Hom(V)} is defined as the group consisting of all
homeomorphisms acting on V.

\noindent\textbf{Definition 3.2.6:} \ If V is a $\ C^{0}$ - space -
time then \emph{K(V)} is defined as the set of all K-causal maps
from V to V.

\noindent \textbf{Proposition 3.2.11 :} \ K(V) is a submonoid of Hom(V).

\noindent \textbf{Proof :} If $\ f_{1}, f_{2}, f_{3} \in K(V) $ then  $\
f_{1} \circ f_{2} \in K(V)$. Also,  $\ f_{1} \circ ( f_{2} \circ
f_{3}) = (f_{1} \circ  f_{2})
 \circ f_{3}$  and identity homeomorphism exists. Hence K(V) is a
submonoid of Hom(V).

\noindent It is obvious that K(V) is a bigger class than the class of
K-conformal maps.

\noindent Thus in a $\ C^{0}$ globally hyperbolic space-time,
every K - causal map $\ f $ where $\ f^{-1} $ is also order
preserving is a causal relation and causal group is the group of
all such mapping which we called K - conformal groups.

\noindent In the light of the definition of quasiorder given in
section 2, we observe that causal cones and K - causal cones fall in
this category, since causal relation `$\ < $ ' and  K - causal
relation `$\ \prec $ ' are reflexive and transitive. If we replace
quasi-order by a causal relation (or K-causal relation), then we
see that an order preserving map is  nothing but a causal map.
Thus an order preserving map is a generalization  of a causal map
(or K-causal map).  These concepts also appear in a branch of
theoretical computer science called domain theory. Martin and
Panangaden  [12] and  S. Janardhan and Saraykar  [10]
have used these concepts in an abstract setting and proved some
interesting results in causal structure of space times. They
proved that order gives rise to a topological structure.

\noindent As far as the \emph{causal topology} on $\ R^{n}$ is
concerned, it is defined as the topology generated by the
fundamental system of neighbourhoods containing open ordered sets\\
$\ M(a,b)$ defined for any $\ a, b \in R^{n} $  with $\ b - a \in
int C $ as : $\ M(a,b) = \{ y \in R^{n} / b - y, \ y - a \in
int C \} $. Gheorghe and Mihul [29] describe  \emph{`causal
topology'} on $\ R^{n}$ and prove that the causal topology of $\
R^{n}$ is equivalent to the Euclidean topology. Causal group $\ G$ is
thus comparable to conformal group of space-time under
consideration. Further any $\ f \in G $ is a homeomorphism in
causal topology and hence it is a homeomorphism in Euclidean
topology.

\noindent If C is a Minkowski cone as discussed in the above
example, then Zeeman [40] has proved that $\ G$ is generated by
translations, dilations and orthochronous Lorentz transformations
of Minkowski
space $\ R^{n} \ (n = 4)$. \\
We can say more for the causal group $\ G$ of Minkowski space.
\\ Let $\ G_{0} = \{ f \in G/ f(0)= 0\} $ .\\
Then $\ G_{0}$ contains the identity homeomorphism. Gheorghe and
Mihul [29]  proved that $\ G$ is generated by the translations of $\
R^{n} $ and by linear transformation belonging to $\ G_{0}$. Hence
$\ G$ is a subgroup of the affine group of $\ R^{n}$. This is the main
result of [29].

\noindent Let $\  G^{'}_{0} =  G_{0} \cap SL(n, R)$. Then $\
G^{'}_{0}$ is the orthochronus Lorentz group under the norm $\
\parallel y
\parallel = [\displaystyle\sum_{i=1}^{q} \mid y^{i} \mid^{2}]^{\frac{1}{2}}$
 for $\ y \in R^{q}, y = (y^{1}, y^{2},\cdots,y^{q}) $.\\
\hspace{10mm} For $\ \parallel y \parallel =
[\displaystyle\sum_{i=1}^{q} \mid y^{i}
\mid^{\alpha}]^{\frac{1}{\alpha}}$,  $\ \alpha > 2 , \ G^{'}_{0}$
is the discrete group of permutations and the symmetries relative
to the origin of the basis vectors of $\ R^{q}$. The factor group
$\ G_{0}/G_{0}^{'}$ is the dilation group of $\ R^{n}$. Also, $\ G$ is
the semi-direct product of the translation group with the subgroup
$\  G^{'}_{0}$  of SL(n,R). Moreover $\  G^{'}_{0}$ is a
topological subgroup of SL(n,R). Similar results have been proved by Borchers and
Hegerfeldt [41]. Thus we have,

\noindent\textbf{Theorem 3.2.12:} \ Let $\ M $ denote n-dimensional
Minkowski space, $\ n \geq 3$ and let $\ T $ be a 1 - 1 map of $\ M $
onto $\ M $. Then $\ T $ and $\ T^{-1} $ preserve the relation $\ (x -
y)^{2} > 0 $ if and only if they preserve the relation $\ (x -
y)^{2} = 0 $. The group of all
such maps is generated by \\
(i)  The full Lorentz group (including time reversal)\\
(ii)Translations of $\ M $ \\
(iii) Dilations ( multiplication by a
scalar)
\\ In our terminology, $\ T $ is a causal map. \\ In the same
paper [41], the following theorem is also proved.

\noindent\textbf{Theorem 3.2.13:} \ Let $\ dim M \geq 3$, and let $\ T $
be a 1 - 1 map of $\ M $ onto $\ M $, which maps light like lines
onto (arbitrary) straight lines. Then $\ T $ is linear.

\noindent This implies that constancy of light velocity c alone implies the
Poincare group upto dilations.

\noindent Thus, for Minkowski space, things are much simpler. For a
space-time of general relativity (a Lorentz manifold) these
notions take a more complicated form where partial orders are $\
J^{+} $ or $\ K^{+}$ .

\section{Causal Orientations and order theoretic approach to Global Hyperbolicity }

\subsection{ \emph{Causal Orientations}}

In this section, we discuss briefly the concepts of \emph{Causal orientations ,
causal structures and causal intervals} which lead to the
definition of a \emph{`Globally hyperbolic homogeneous space'}.\\
These notions cover Minkowski Space and homogeneous cosmological models in general relativity.
We also discuss domain theoretic approach to causal structure of space-time and comment
on the parallel concepts appearing in these approaches.

\noindent Let $\ M $ be a $\ C^{1}$ (respectively smooth) space-time.
For $\ m \in M,  \ T_{m}(M) $ denotes the tangent space of $\ M $ at m \ and T($\ M $) denotes
the tangent bundle of $\ M $. The derivative of a differentiable
map $\ f : M \rightarrow N $ at m will be denoted by $\ d_{m}f :
T_{m}M \rightarrow T_{f(m)}N $. A $\ C^{1} $ (respectively smooth)
causal structure on $\ M $ is a map which assigns to each point $\
m \in M $ a nontrivial closed convex cone C(m) in $\ T_{m}M $ and
it is $\ C^{1}$(smooth) in the following sense:

\noindent We can find an open covering $\ \{ U_{i} \}_{i \in I} $ of $\ M $,
smooth maps $\ \phi_{i}: U_{i} \times R^{n} \rightarrow T(M) $with
$\ \phi_{i}(m,M) \in T_{m}(M)$ and a cone C in $\ R^{n} $ such
that $\ C(m)  = \phi_{i} (m,C)$ .

\noindent The causal structure is called \emph{generating} (respectively
proper, regular) if C(m) is generating (proper, regular) for all
m. A map $\ f : M \rightarrow M $ is called  \emph{causal} if $\
d_{m}f(C(m)) \subset C(f(m)) $ for all $\ m \in M $. These
definitions are obeyed by causal structure $\ J^{+}$ in a causally
simple space-time and causal maps of Garc$\acute{i}$a-Parrado and
Senovilla [32]. If we consider $\ C^{0}$- Lorentzian manifold with
a $\ C^{1}$ -metric so that we can define null cones, then
these definitions are also satisfied by causal structure $\ K^{+}$
and K-causal maps. Thus the notions
defined above are more general than those occurring in general
relativity at least in a special class of space-times.   Rainer
[42] called such a causal structure an   \emph{ultra weak cone
structure}  on $\ M $ where $\ m \in \ int M  $.

\noindent We now define $\ G$- invariant causal structures where $\ G$ is a Lie group and discuss
some properties of such structures. If a Lie group $\ G$ acts smoothly on
$\ M $ via $\ (g,m) \mapsto g.m.$, we denote the diffeomorphism $\ m \mapsto g.m $ by $\
l_{g}$.

\noindent\textbf{Definition 4.1.1:} Let $\ M $ be a manifold with a
causal structure and $\ G$ a Lie group acting on $\ M $. Then the
causal structure is called \emph{$\ G$ - invariant} if all $\ l_{g}, \
g \in G $, are causal maps. If H is a Lie subgroup of $\ G$ and $\ M =
G/H $ is homogeneous then a $\ G$-invariant causal structure is
determined completely by the cone $\ C = C(0) \subset T_{o}M $,
where $\ o = {H} \in G/H $. Moreover C is proper, generating etc
if and only if this holds for the causal structure. We also note
that C is invariant under the action of H on $\ T_{o}(M)$ given by
$\ h \mapsto d_{0}l_{h}$. On the other hand, if $\ C \in Cone_{H}
(T_{o}(M))$,then we can define a field of cones by $\ M
\rightarrow T_{\alpha .0}(M) : \ aH \mapsto C(\alpha H) = d_{0} l_{a}(C)$.

\noindent This cone field is $\ G$-invariant, regular and satisfies $\ C(0) = C $.
Moreover the mapping $\ m \mapsto C(m)$ is also smooth in the
sense described above. If this mapping is only continuous in the
topological sense,  for all m  in M,  then Rainer [42] calls such
cone structure, a \emph{weak local cone structure} on $\ M
$.\\
We have the following theorem.

\noindent\textbf{Theorem 4.1.1:} Let $\ M = G/H $ be homogeneous.
Then $\ C \mapsto (\alpha H \mapsto d_{0}l_{a}(C)) $ defines a
bijection between $\ Cone_{H}(T_{o}(M)) $ and the set of
$\ G$-invariant, regular causal structures on $\ M $.

\noindent We call  a mapping $\ \nu : [a,b] \rightarrow M $ as
\emph{absolutely continuous }if for any coordinate chart $\ \phi :
U \rightarrow R^{n}$, the curve $\ \eta = \phi \circ \nu :
\nu^{-1}(U) \rightarrow R^{n} $ has absolutely continuous
coordinate functions and the derivatives of these functions are
locally bounded.

\noindent Further, we define a \emph{C-causal curve}: \ Let $\ M = G/H $ and
$\ C \in Cone_{G}(T_{o}M)$. An absolutely continuous curve $\ \nu
: [a,b] \rightarrow M $ is called C - causal ( \emph{Cone causal
or conal}) if $\ \nu^{'}(t) \in C(\nu(t)) $ whenever the
derivative exists.

\noindent Next, we define a relation `$\ \leq_{s}$'  (s for strict) of $\ M
$ by $\ m \leq_{s} n $ if there exists a C-causal curve $\ \nu $
connecting m with n. This relation is obviously reflexive and
transitive. Such relations are called \emph{causal orientations}
or \emph{quasi - orders}. They give rise to causal cones as we saw in section 2.

\noindent\textbf{Note :} A reader who is familiar with the books
by Penrose [30], Hawking and Ellis [4] or Joshi [6] will
immediately note that the above relation is our familiar causal
order $\ J^{\pm}$ in the case when $\ M $ is a space-time in
general relativity.

\noindent We ask the question : Which of the space-times $\ M $
can be written as $\ G/H $?  G\"{o}del universe, Taub universe and
Bianchi universe are some examples of such space-times. They are
all spatially homogeneous cosmological models. Isometry group of a
spatially homogeneous cosmological model may or may not be
abelian. If it is abelian, then these are of Bianchi type I, under
Bianchi classification of homogeneous cosmological models. Thus
above discussion applies to such models.

\noindent As an example to illustrate above ideas, we again consider a
finite dimensional vector space $\ M $ and let C be a closed
convex cone in $\ M $. Then we define a causal Aut(C) - invariant
orientation on $\ M $ by   $\ u \leq v $ iff $\  v - u \in C $ .
Then `$\ \leq $' is antisymmetric iff C is proper. In particular
$\ H^{+}(n,\emph{R}) $ defines a $\ GL(n,\emph{R})$ -invariant
global ordering in H(n,\emph{R}). Here H(n,R) are $\ n\times n $
real orthogonal matrices (Hermitian if \emph{R} is replaced by
\emph{C}) and  $\ H^{+}(n,R) = \{ X \in H( m,\emph{R}) / X $ is
positive definite $\}$ is an open convex cone in H(n,R). (the
closure of $\ H^{+}(n,R) $ is the closed convex cone of all
positive semi definite matrices in H(n,R)). Also, the light cone
$\ C \subset R^{n+1 }$ defines a $\ SO_{O}(n,1)$ -invariant
ordering in $\ R^{n+1}$. The space $\ R^{n+1}$ together with this
global ordering is  the (n+1)-dimensional Minkowski space.

\noindent Going back to the general situation we note that in general, the
graph \\ $\ M_{\leq_{s}} = \{(m,n) \in M \times M / m \leq_{s} n
\} $ of `$\ \leq_{s}$' is not closed in $\ M \times M $. However,
if we define $\ m \leq n \Leftrightarrow (m,n) \in
\overline{M}_{\leq_{s}}$, then it turns out that `$\ \leq $' is a
causal orientation. This can be seen as follows:\\
`$\ \leq $'  is obviously reflexive. We show that it is
transitive: \\ Suppose $\ m \leq n \leq p $ and let $\
m_{k},n_{k},n_{k}^{'},p_{k} $ be sequences such that $\ m_{k}
\leq_{s} n_{k}, \\ n_{k}^{'} \leq_{s} p_{k}, \ m_{k} \rightarrow \
m,   \ n_{k} \  \rightarrow \ n, \  n_{k}^{'} \ \rightarrow \ n $
and $\ p_{k} \rightarrow  \ p $. Now we can find a sequence $\
g_{k} $ in $\ G$ converging to the identity such that $\ n_{k}^{'} =
g_{k}n_{k}$. Thus $\ g_{k}m_{k} \rightarrow  m $ and $\ g_{k}n_{k}
\leq_{s} p_{k} $ implies $\ m \leq p $.

\noindent The above result resembles the way in which $\ K^{+} $ was
constructed from $\ I^{+}$.\\
The following definitions are analogous to $\ I^{\pm},J^{\pm}$ or
$\ K^{\pm}$ and so is the definition of interval as $\ I^{+}(p)
\cap I^{-}(q) ( J^{+}(p) \cap J^{-}(q)$  or $\ K^{+}(p) \cap
K^{-}(q))$: \\
Given any causal orientation `$\ \leq $' on $\ M $, we define for
$\ A \subset M$,

\noindent $\  \uparrow A = \{ y \in M / \exists a \in A \ $ with $\ a \leq y
\}$   and

\noindent $\  \downarrow A = \{ y \in M / \exists a \in A \ $ with $\ y \leq
a
\}$.\\
Also, we write $\ \uparrow x = \uparrow \{x\} $ and $\ \downarrow
x = \downarrow \{x\} $. \\
The intervals with respect to this causal orientation are defined
as

\noindent $\ [m,n]_{\leq}= \{z \in M / m \leq z \leq n \} = \uparrow m \
\cap
\downarrow n $ .\\
Finally we introduce some more definitions.

\noindent\textbf{Definitions 4.1.2:} Let $\ M $ be a space-time.\\
(1) a causal orientation `$\ \leq $' on $\ M $ is called
\emph{topological} if its graph $\ M_{\leq} $ in $\ M \times M $
is closed. \\
(2) a space $\ (M,\leq )$  with a topological causal orientation
is called a \emph{causal space}. If `$\ \leq $' is, in addition,
antisymmetric, that is a partial order, then $\ (M,\leq )$ is
called \emph{globally ordered} or \emph{ordered}.\\
(3) Let $\ (M, \leq )$ and $\ (N,\leq )$ be two causal spaces and
let $\ f:  M \rightarrow  N $ be continuous. Then $\ f $ is called
\emph{order preserving} or \emph{monotone} if $\ m_{1} \leq  m_{2}
 \Rightarrow f(m_{1}) \leq f(m_{2})$. \\
(4) Let $\ G$ be a group acting on $\ M $. Then a causal orientation
$\ \leq $ is called \emph{$\ G$-invariant} if $\ m \leq n  \Rightarrow
a.m \leq  a.n , \  \forall \ a \in G $.\\
(5) A triple $\ ( M, \leq, G)$ is called a \emph{Causal
$\ G$-Manifold} or \emph{causal} if `$\ \leq $' is a topological
$\ G$-invariant causal orientation.
\\ Thus referring to partial order $\ K^{+}$, we see, in the light of
above definitions (1) and (2), that $\ \leq_{K}$ is topological
and $\ ( M, \leq_{K})$ is a causal space. A K-causal map satisfies
definition (3).

\noindent For a homogeneous space $\ M = G/H $ carrying a causal
orientation such that $\ ( M, \leq, G)$ is causal, the intervals
are always closed subsets of $\ M $. If the intervals are compact,
we say that $\ M = G/H $ is \emph{globally hyperbolic}. We use the
same definition for a space-time where intervals are $\
J^{+}(p)\cap J^{-}(q)$. Thus globally hyperbolic space-times can
be defined by using causal orientations for homogeneous spaces. In
this setting, intervals are always closed, as in causally
continuous space-times.

\subsection{ \emph{Domain Theory and Causal Structure}}

As the last part of our article, we discuss the central concepts and definitions
of domain theory, as we observe that these concepts  are related to causal
structure of space-time  and also to space-time topologies.

\noindent The relations $\ < and \ll $ discussed in section 3 have been generalised
to abstract orderings using the concepts in Domain Theory and also many
interesting results have been proved related to causal structures of space-time in
general relativity. For definitions and preliminary results in domain theory,
we follow Abramsky and Jung [43] and Martin and Panangaden [13].
We have expanded some of the proofs which follow in this section, as it gives a better
understanding of these concepts and their applications.

\noindent\textbf{Definition 4.2.1: } A \emph{poset} is defined as a
partially ordered set, i.e. a set together with a reflexive, anti-
symmetric and transitive relation.

\noindent Domain theory deals with partially ordered sets to model a domain
of computation and the elements of such an order are interpreted
as pieces of information or results of a computation where
elements of higher order extend the information of the elements
below them in a consistent way.

\noindent\textbf{Definition 4.2.2: } Let $\ (P, \sqsubseteq)$ be a
partially ordered set. An  \emph{upper bound}  of a subset S of a
poset P is an element b of P, such that $\ x \sqsubseteq  b , \
\forall x \in \ S $. The dual notion is called \emph{lower bound}.

\noindent A concept that plays an important role in domain theory is the one
of a directed subset of a domain, i.e. of a non-empty subset  in
which each two elements have an upper bound.

\noindent\textbf{Definition 4.2.3: }A nonempty subset $\ S \subseteq P$
is \emph{directed} if for every $\ x, y $ in S, $\ \exists \ z \in
S \  \ni: \ x , y \sqsubseteq \ z $. The supremum of $\ S
\subseteq P$ is the least of all its upper bounds provided it
exists and is denoted by $\ \bigsqcup S $.

\noindent This means that every two pieces of information with in the
directed subset are consistently extended by some other element in
the subset.

\noindent A nonempty subset $\ S \subseteq P$ is \emph{filtered } if for
every $\ x $, y in S ,$\ \exists z \in S \ni: \ z  \sqsubseteq \ x
, y $. The infimum of $\ S \subseteq P$ is the greatest of all its
lower bounds provided it exists and is denoted by $\ \bigwedge S
$.

\noindent In the partially ordered set $\ ( R , \leq ) $ where $\
R $ is the set of real numbers and $\ \leq $ denotes the relation
\emph{less than or equal to}, the subset [0 , 1] is  directed
with supremum 1 and is  filtered  with infimum 0.

\noindent\textbf{Remark }: \\ (i) $\ \forall \ x \in P $ , $\ \{x
\} $ is a directed set. \\ (ii) \ In the theory of metric spaces,
sequences play a role that is analogous to the role of directed
sets in domain theory in many aspects. \\
(iii) \ In the formalization of order theory, \emph{limit} of a
directed set is just the \emph{least upper bound} of the directed
set. As in the case of limits of sequences,least upper bounds of
directed sets do not always exist.

\noindent The domain  in which all consistent specifications converge is of
special interest and is defined as follows:

\noindent\textbf{Definition 4.2.4:} A \emph{dcpo}(directed complete
partial order) $P$ is a poset in which every directed subset has a
supremum.

\noindent The poset $\ ( R , \leq ) $ is not a dcpo, as the
directed subset $\ ( 0 , \infty ) $ does not have a supremum.

\noindent Using partial order, some topologies can be derived. For example,

\noindent\textbf{Definition 4.2.5:} A subset U of a poset $P$ is
 \emph{Scott open} if \\
(i) U is an upper set: i.e. $\ x \in U $ and $\  x \sqsubseteq y
 \Rightarrow  y \in U, $ and \\
(ii) U is inaccessible by directed suprema: i.e. for every
directed $\ S \subseteq P $ with a supremum, $\ \bigsqcup S \in U
\Rightarrow  S \cap U \neq \phi $.

\noindent \ The collection of all Scott open sets on $P$ is called
the \emph{Scott topology}.

\noindent For the poset $\ ( R , \leq ) , \ ( 1 ,
\infty ) $ is Scott open.

\noindent A more elaborate approach leads to the definition of
order of \emph{approximation} denoted by $\ `\ll '$ which is also
called the \emph{way - below relation}.

\noindent\textbf{Definition 4.2.6:}  For elements $\ x , y $ of a
poset, $\ x \ll y $  iff for all directed sets S with a supremum,
$\ y  \ \sqsubseteq  \bigsqcup S \ \Rightarrow \exists \ s \in S
\ni: \ x
\sqsubseteq s $. \\
Define, $\ \Downarrow x = \{ a \in  P / a \ll x  \}$  and $\
\Uparrow x = \{ a \in  P /  x \ll a \}$ .

\noindent In an ordering of  sets, an infinite set is way above any of its
finite subsets. On the other hand, consider the directed set of
finite sets $\ \{0\}, \{0,1\}, \{0,1,2\} ...$ . The supremum of
this set is the set N of all natural numbers. i.e , no infinite
set is way below N.

\noindent\textbf{Definition 4.2.7:}  An element $\ x  $ in a poset $\ P $
is said to be  \emph{compact} if $\ x \ll x $.

\noindent\textbf{Proposition 4.2.1: }: $\ x \ll y \Rightarrow x \sqsubseteq y
$.

\noindent\textbf{Proof :} Let $\ x \ll y $. Consider the directed
set $\ \{y \}$. Since $\ \bigsqcup  \{y \} = y $, by definition $\
x \sqsubseteq y $.

\noindent\textbf{Proposition 4.2.2:}: The relation $\ `\ll ' $ is not
necessarily reflexive.

\noindent\textbf{Proof :} Let S be a directed set with $\ x  =
\bigsqcup S $ and  $\ x $ is not in S. Then $\ x \sqsubseteq s , \
s \in S $ is false.

\noindent\textbf{Proposition 4.2.3: }: $\ x \sqsubseteq y \ll \ z \Rightarrow
\ x \ll z $.

\noindent\textbf{Proof :} Let S be a directed set with $\ z
\sqsubseteq \bigsqcup S $. Now $\ y \ll z \ \Rightarrow \exists \
s \in S $ such that $\ y \sqsubseteq s $. Since $\ x \sqsubseteq y
$ , we have $\ x  \sqsubseteq s $. This holds for each directed
set with
$\ \bigsqcup S \sqsupseteq z $. Hence $\ x \ll z$.\\
Let $\ x \ll y \sqsubseteq z $. If S is a directed set with $\ z
\sqsubseteq \bigsqcup S $ then $\ y \sqsubseteq \bigsqcup S $.
Hence $\ x \ll y \Rightarrow \exists \ s \in S $ such that $\ x
\sqsubseteq s $. Thus $\ x \ll z $.

\noindent\textbf{Definition 4.2.8:} For a subset X of a poset $P$, define
\\ $\ \uparrow X := \{y \in P / \exists \ x \in X,  x
\sqsubseteq y \}$  and \\
$\ \downarrow X :=  \{y \in P / \exists \ x \in X,  y \sqsubseteq
x \}$ \\
Then, $\ \uparrow x =\uparrow \{x \}$ and $\ \downarrow x =
\downarrow \{x \} $ for $\ x \in X $.

\noindent In $\ ( R , \leq ) $, $\ \uparrow \{x \} = [ x , \infty
)$ and $\ \downarrow \{x \} =  ( - \infty , x ]$

\noindent A subset of elements which is sufficient for getting all other
elements as least upper bounds can be defined as follows:

\noindent\textbf{Definition 4.2.9:} A \emph{basis} for a poset $P$ is a
subset B such that $\ B \cap \Downarrow x $ contains a directed
set with supremum $\ x $ for all $\ x $ in  $P$.

\noindent A poset is  \emph{continuous} if it has a basis. A poset
is $\ \omega $-\emph{continuous} if it has a countable basis.

\noindent Continuous posets have an important property that they
are interpolative.

\noindent\textbf{Proposition 4.2.4:} $\ \Downarrow x $ is a directed set in
a continuous poset $P$.

\noindent\textbf{Proof :}  Let B be a basis in $P$. Then $\ B \cap
\Downarrow x $ is a directed set with  $\ x = \bigsqcup S $.
Let $\ y, z \in \Downarrow x $. Then $\ y \ll x $ and $\ z \ll x
$. Now $\ y \ll x $ implies $\ \exists s_{1} \in S $ such that $\
y \sqsubseteq s_{1} $. \ $\ z \ll x $ implies  $\ \exists s_{2}
\in S $ such that $\ z \sqsubseteq s_{2} $. Now both $\ s_{1} ,
s_{2} \in S $ and S is directed. Therefore, $\ \exists \ s \in S $
such that $\ s_{1} , s_{2} \sqsubseteq s $. Hence $\ s \in
\Downarrow x $ and $\ y \sqsubseteq s , z \sqsubseteq s $. Thus $\
\Downarrow x $ is a directed set.

\noindent\textbf{Proposition 4.2.5:} $\ \bigsqcup \Downarrow x = x $, in a
continuous poset $P$.

\noindent\textbf{Proof:} For every $\ y \in \Downarrow x , \ y \ll
x $. Therefore, $\ y \sqsubseteq x $. i.e. $\ x  $ is an upper
bound of $\ \Downarrow x $. Let a be any other upper bound of $\
\Downarrow x $. Since $P$ is a continuous poset, $\ B \cap
\Downarrow x $ contains a directed set S with $\ \bigsqcup S  = x
$. Obviously, $\ S \subseteq \Downarrow x$. Hence $\ \bigsqcup S
\sqsubseteq \bigsqcup \Downarrow x \sqsubseteq $ any upper bound
of $\ \Downarrow x $.  Therefore, $\ x \sqsubseteq a $. Thus $\ x
= \bigsqcup \Downarrow x $ where $P$ is a continuous poset .

\noindent\textbf{Proposition 4.2.6:} If  $\ x \ll y $ in a continuous
poset $P$, then there is $\ z \in P$  with $\ x \ll z \ll y $ (that
is, continuous posets are interpolative ). Actually a more general
result is true namely, if G is a finite subset of $P$ with $\ G \ll
y $, i.e. $\ \forall \ x \in G, x \ll y $, then $\ \exists \ z \in
P $ such that $\ G \ll z \ll y $.

\noindent\textbf{Proof :}\ Let $\ A = \{ a \in P  /  \ \exists \
a^{'} \in P$ with $\ a \ll a^{'} \ll y \} $. We claim that A is
non - empty. Consider $\ x \in M , x \ll y $. Now $\ B \cap
\Downarrow x $ contains a directed set S with $\ \bigsqcup S = x
$. Let $\ a \in S $. Then $\ a \in \Downarrow x $. Therefore, $\ a
\ll x $ and $\ x \ll y $. Hence $\ a \in A $.

\noindent Now we claim that A is a directed set. Let $\ a, b \in A $. Then
$\ \exists a^{'} , b^{'} \in P $ such that $\ a \ll a^{'} \ll y $
and $\ b \ll b^{'} \ll y $. Since $\ a^{'} , b^{'} \in \Downarrow
y $ and $\ \Downarrow y $ is a directed set, $\ \exists \ c^{'}
\in \Downarrow y $ such that $\ a^{'} , b^{'} \sqsubseteq c^{'},
c^{'} \ll y $. Using directedness of $\ \Downarrow c^{'} $, we
have, $\ a \ll a^{'} \sqsubseteq c^{'} , \ b \ll b^{'} \sqsubseteq
c^{'} $. Therefore, $\ a, b \in \Downarrow c^{'} $ and hence, $\
\exists \ c \in \Downarrow c^{'} \ni: a, b \sqsubseteq c $.

\noindent As $\ c \ll c^{'} $ and $\ c^{'} \ll y $ , we have $\ c \in
\Downarrow y $. Thus, given $\ a, b \in A, \exists \ c \in A \ni:
a, b \sqsubseteq c  $. Hence A is a directed set.

\noindent We now show that $\ y = \bigsqcup \Downarrow y = \bigsqcup A $.
Let $\ y^{'} \ll y $. Then for each $\ r \in \Downarrow y^{'}, \\
r \ll y^{'} \ll y $.  Therefore, $\ r \in A$ which implies $\
\Downarrow y^{'} \subseteq A $. Hence $\ \bigsqcup \Downarrow
y^{'} \sqsubseteq  \bigsqcup A $. i.e., $\ y^{'} =  \bigsqcup
\Downarrow y^{'} \sqsubseteq \bigsqcup A $. This holds holds for
each $\ y^{'} \ll y $. Since $\ B \cap \Downarrow y $ contains a
directed set S with $\ \bigsqcup S = y $, for each $\ y^{'} \in S
, y^{'} \ll y $. Therefore, $\ y^{'} \sqsubseteq   \bigsqcup A $.
Hence $\ \bigsqcup S \sqsubseteq \bigsqcup A$. But $\ \bigsqcup S
= y $. Therefore, $\ y \sqsubseteq \bigsqcup A $. But by
definition, each element of A is below y. Therefore, $\ \bigsqcup
A \sqsubseteq y $. Hence $\ y = \bigsqcup A $.

\noindent For each $\ x \in G , x \ll y = \bigsqcup A $, and A is a directed
set. So, by definition of $\ '\ll', \exists \ z_{x} \in A, \ni: x
\sqsubseteq z_{x} $. Since G is finite, $\ z_{x} $ in A are finite
in number. So, by directedness of A, $\ \exists \ z^{'} \in A \ni:
x \sqsubseteq z^{'} \ \forall \ x \in G. $ Now $\ z^{'} \in A
\Rightarrow \exists \ z \ni: z^{'} \ll z \ll y$. Therefore, $\ x
\ll z \ll y,  \  \forall \ x \in G $. i.e. $\ G \ll z \ll y $.

\noindent Then we have,

\noindent\textbf{Theorem 4.2.7:} The collection $\ \{ \Uparrow x / x
\in P \}$ is a basis for the Scott topology on a continuous poset.

\noindent\textbf{Proof: } We first show that $\ \Uparrow x $ is
Scott open for each $\ x $ in $P$. Let $\ y \in \Uparrow x , \   y
\sqsubseteq z $. Then $\ x \ll y \sqsubseteq z $. So, we have $\ x
\ll z $ and hence $\ z \in \Uparrow x $. Thus $\ \Uparrow x $ is
an Upper set. Let S be any directed set with a supremum such that
$\ \bigsqcup S \in \Uparrow x$. Let $\ y = \bigsqcup S $. Thus $\
y \gg x $. By interpolativeness of `$\ \ll ' , \  \exists \ z \in
P \ \ni: \  y \gg z \gg x , \ z \ll y \ $ and $\  \ y = \bigsqcup
S $.

\noindent Therefore $\ \exists \ s \in S \ni: z \sqsubseteq s $. Then $\ x
\ll z \sqsubseteq s$ and hence $\ x \ll s $. Further $\ s \in S $.
So, $\ s \in \Uparrow x $ and $\ s \in S $. Therefore, $\ S \cap
\Uparrow x \neq \phi $. Thus $\ \Uparrow x $ is Scott open for each $\ x \in P $. Let $\ x
\in P $, and \emph{U} be a Scott open set with $\ x \in U $.
Consider $\ B \cap \Uparrow x $. It contains a directed set say S
with $\ \bigsqcup S = x $. Since $\ x \in U $, it follows that $\
S \cap U \neq \phi $. Let $\ y \in S \cap U $. Obviously $\ x \in
\Uparrow y $. Let $\ z \in \Uparrow y $. Since $\ y \in U $ and $\
y \sqsubseteq z $ we must have$\ z \in U $. Thus for each $\ x \in
P $ and Scott open set \emph{U} , $\ x \in \ U , \exists \ y \ni:
\ x \in \ \Uparrow y  $ and $\ \Uparrow x \subseteq U $. Hence, $\ \{
\Uparrow x / \ x \in P \} $ forms a basis for the Scott topology.

\noindent Lawson topology can be defined as,

\noindent\textbf{Definition 4.2.10:} \ The \emph{Lawson topology} on a
continuous poset $P$ has as a basis all sets of the form $\ \Uparrow
x \sim \uparrow F $, for $\ F \subseteq P $ finite.

\noindent\textbf{Definition 4.2.11: } A continuous poset $P$ is
 \emph{bicontinuous} if for all $\ x , y $ in  $P$ \\
$\ x  \ll y $ iff for all filtered $\ S \subseteq P $ with an
infimum, $\ \bigwedge S \sqsubseteq x \Rightarrow \exists \ s \in
\ S \ni:  \ s \sqsubseteq \ y $ and   for each $\ x \in P $, the
set $\ \Uparrow x $ is filtered with infimum $\ x $.

\noindent\textbf{Definition 4.2.12: } A \emph{domain} is a continuous
poset which is also a \emph{dcpo}.

\noindent\textbf{Proposition 4.2.8:} On a bicontinuous poset $P$, sets of the
form
\\ $\ (a, b) := \{ x \in P / a \ll x \ll b \} $  form a basis for a
topology. This topology is called the  \emph{interval topology}.

\noindent\textbf{Proof :} For any $\ x \in P , \Uparrow x $ is
filtered with infimum $\ x $ and $\ \Downarrow x $ is directed
with supremum $\ x $. Due to bicontinuity, $\ \Uparrow x, \
\Downarrow x $ are non- empty. Let $\ a \in \Downarrow x, \ b \in
\Uparrow x $. Then $\ a \ll x \ll b $. \\ Let $\ x \in P $ be such
that $\ a \ll x \ll b $ and $\ a_{1} \ll x \ll b_{1} $. Then $\ a,
a_{1} \in \Downarrow x $. Since $\ \Downarrow x $ is a directed
set, $\ \exists \ a_{2}  \in \Downarrow x \ni: a, a_{1}
\sqsubseteq a_{2} $. \\ Similarly, $\ b, b_{1} \in \Uparrow x $ \
which is filtered. Therefore, $\ \exists \ b_{2} \in \ \Uparrow x
$ such that \ $\ b_{2} \sqsubseteq b, b_{1} $. Obviously, $\ a_{2}
\ll x \ll b_{2} $. Further, if y is such that $\ a_{2} \ll y \ll
b_{2} $, then $\ a \sqsubseteq a_{2} \ll y $ and $\ y \ll b_{2}
\sqsubseteq b  \ \Rightarrow a \ll y \ll b $. i.e., $\ y \in a \ll
... \ll b $. Similarly, $\ y \in a_{1} \ll ... \ll b_{1} $. Hence
$\ ( a, b ) $ forms a topology on $P$. \\We recall some more definitions regarding causal
structure of space-time and elaborate and modify proofs of certain
theorems regarding causality conditions.

\noindent\textbf{Definition 4.2.13:} \ The relation $\ J^{+} $ is
 defined as   $\ p \sqsubseteq q \equiv q \in J^{+}(p) $.

\noindent\textbf{Proposition 4.2.9:}  Let $\ p, q, r \in M $. Then \\
(i) The sets $\ I^{+}(p) $ and $\ I^{-}(p) $ are open. \\
(ii) $\ p \sqsubseteq q $ and $\ r \in I^{+}(q) \Rightarrow  r \in
I^{+}(p)$ \\
(iii)$\  q \in  I^{+}(p)$ and $\ q \sqsubseteq r \Rightarrow r \in
I^{+}(p)$ \\
(iv) $\ \overline{I^{+}(p)} = \overline{J^{+}(p)}$ and $\
\overline{I^{-}(p)} = \overline{J^{-}(p)}$.

\noindent We assume strong causality which can be characterized as follows:

\noindent\textbf{Theorem 4.2.10:} A space-time $\ M $ is strongly causal
iff its Alexandrov topology is Hausdorff iff its Alexandrov
topology is the manifold topology.

\noindent\textbf{Definition 4.2.14:} A space-time $\ M $ is
\emph{globally hyperbolic } if it is strongly causal and if $\
\uparrow \ a \ \cap \downarrow \ b $ is compact in the manifold
topology, for all $\ a , b $ in $\ M $.

\noindent\textbf{Lemma 4.2.11: } If $\ (x_{n}) $ is a sequence in a
globally hyperbolic space-time $\ M $ with $\
x_{n} \sqsubseteq  x $ for all n, then \\
$\ \displaystyle\lim_{n \rightarrow \infty} x_{n} = x \Rightarrow
\displaystyle\bigsqcup_{n \geq 1} x_{n} = x $.

\noindent\textbf{Lemma 4.2.12:} For any $\ x \in M ,  I^{-}(x)$ contains
an increasing sequence with supremum $\ x $.

\noindent\textbf{Proposition 4.2.13:} In a globally hyperbolic
space-time M, $\ x \ll y  \Leftrightarrow y \in I^{+}(x)$  for all
$\ x , y $ in $\ M $. Here $M$ is a bicontinuous poset.

\noindent\textbf{Proof :} \ Let $\ x \ll y $. Then, there
is an increasing sequence $\ (y_{n})$ in $\ I^{-}(y)$ with $\ y =
\bigsqcup y_{n}$. Since $\ x \ll y $, there exists $\ n $ such
that $\ x \sqsubseteq y_{n}$. \\ Hence, $\ x
\sqsubseteq y_{n} $ and  $\ y_{n} \in I^{-}(y) \Rightarrow \ x \in
\ I^{-}(y)$. That is, $\ y \in I^{+}(x)$.

\noindent Let $\ y \in I^{+}(x)$. To prove $\ x \ll y $, we have to prove
that if S is any directed set with $\ y \sqsubseteq \bigsqcup S $,
then $\ \exists \ s \in S $ such that $\ x \sqsubseteq s$. Since
$\ y \sqsubseteq \bigsqcup  S, \bigsqcup S \in J^{+} (y) $ , we
have $\ y \in I^{+}(x) $ and hence $\
\bigsqcup S \in I^{+}(x) $.\\
Case 1 :  If $\ \bigsqcup S \in S $ , then we take $\ s =
\bigsqcup S, $ and hence the proof.\\
Case 2 :  Let $\ \bigsqcup S $ is not in S.Then S must be
infinite. Let $\ \bigsqcup S = z $.Consider $\ s_{1} , s_{2} \in S
$. Then we can find $\ s_{3} $ such that $\ s_{1} \sqsubseteq
s_{3} ,s_{2} \sqsubseteq s_{3} $. ( If $\ s_{3} $ coincides with
$\ s_{1}  \ or \ s_{2} $ in that case , we have $\ s_{1}
\sqsubseteq s_{2} \ or \ s_{2} \sqsubseteq s_{1} $ ). Consider
then another element of S different from $\  s_{1} ,s_{2}, s_{3}
$, $\ \exists  s_{4} \in S \ni:  s_{3} \sqsubseteq  s_{4} ... $.
We can proceed in this way to get a strictly increasing sequence
in S. If we denote this set by $\ S^{'} $ , then $\ \bigsqcup
S^{'} = \bigsqcup S  = z $. ( For, $\ \bigsqcup S^{'} = \bigsqcup
S , \ as \ S^{'} \subseteq S $. If $\ \bigsqcup S^{'} \sqsubseteq
\bigsqcup S $ and $\ \bigsqcup S^{'} \neq \bigsqcup S $ then
either there exists  s in S such that $\ \bigsqcup S^{'} $ and s
are not related or $\ \bigsqcup S^{'} \sqsubseteq s $ and  $\
\bigsqcup S^{'} \neq s $. Both are ruled out as S is a directed
set. Thus , $\ S^{'} $ is a strictly increasing chain in S with $\
\bigsqcup S^{'} = \bigsqcup S $.)

\noindent For this $\ S^{'} =  \{ s_{1}, s_{2} ... \}$, we consider compact
sets $\ J^{+} (s_{i}) \cap J^{-}(z) $. Then,  $\ \{ J^{+} (s_{i})
\cap J^{-}(z) \} $ will be a decreasing sequence of compact sets
whose intersection is z which is in the open set $\ I^{+} (x) $.
Hence, for some $\ s_{i}, \ J^{+}(s_{i}) \cap J^{-}(z) \subseteq
I^{+}(x) $. Otherwise, from each of the above compact sets we can
find $\ x_{i} $ such that $\ x_{i} $ is not in $\ I^{+}(x) $,
where $\ x_{i} $ is an increasing sequence with $\ z = \sup\{x_{i}
\} $ and the open set $\ I^{+}(x) $ not intersecting the sequence.
This is not possible. Therefore $\ \ J^{+}(s_{i}) \ \cap \ J^{-}(z) \ \subseteq I^{+}(x)
$ which implies $\ s_{i} \in I^{+}(x) $ as $\ s_{i} \sqsubseteq z
$. i.e. $\ x \sqsubseteq s_{i} $ and hence  $\ x\ll y $.

\noindent The above proof is a modified version of that given in [13].

\noindent\textbf{Theorem 4.2.14:} If  M is  globally hyperbolic then
$\ (M, \sqsubseteq ) $ is a bicontinuous poset with $\ \ll =
I^{+}$ whose interval topology is the manifold topology.

\noindent Causal simplicity  also has a characterization in order-theoretic
terms.

\noindent\textbf{Theorem 4.2.15:} Let   $\ (M, \sqsubseteq) $ be a
continuous poset with $\ \ll = I^{+}$. Then the following are
equivalent: \\
(i) $\ M $ is causally simple. \\
(ii) The Lawson topology on $\ M $ is a subset of the interval
topology on $\ M $.

\noindent We now give definitions and results from a recent article by K. Martin and P. Panangaden [13].

\noindent\textbf{Definition 4.2.15:} Let $\ (X, \leq) $ be a globally hyperbolic poset. A subset $\ \pi \subseteq X $ is
a causal curve if it is compact, connected and linearly ordered. Let $\ \pi(0) = \bot $ and $\ \pi (1) = \top $ where $\ \bot$ and $\ \top $ are the least and greatest elements of $\ \pi $. For $\ P, Q \subseteq X $,\\
$\ C(P,Q) = \{ \pi / \pi \ causal \ curve, \pi(0) \in P,  \pi(1) \in Q \}$ \\
called  the space of causal curves between $\ P $ and $\ Q $.

\noindent It is clear that a subset of a globally hyperbolic space-time $\ M $ is the image of a causal curve iff it is the image of a
continuous monotone increasing $\ \pi : [0, 1] \rightarrow M $ iff it is a compact connected
linearly ordered subset of $\ (M, \leq) $.

\noindent\textbf{Theorem 4.2.16:} If $\ (X, \leq) $ is a separable globally hyperbolic poset,
then the space of causal curves $\ C(P,Q) $ is compact in the Vietoris topology
and hence in the upper topology.

\noindent This result plays an important role in the proofs of certain singularity theorems in [5],
in establishing the existence of maximum length geodesics in [4]
and in the proof of certain positive mass theorems in [45].

\noindent Also, Globally hyperbolic posets are very much like the real line. A well-known domain theoretic
construction pertaining to the real line extends in perfect form to the globally
hyperbolic posets:

\noindent\textbf{Theorem 4.2.17:} The closed intervals of a globally hyperbolic poset $\ X$,
$\ \textbf{IX} = \{ [a,b] / a \leq \ b \ and \ a, \ b \in X \}$ ordered by reverse inclusion
$\ [a, b] \subseteq [c, d] \equiv [c, d] \subseteq [a, b]$ form a continuous domain with
$\ [a, b] \ll [c, d] \equiv a \ll \ c \ and \ d \ll b $.
The poset $\ X $ has a countable basis iff $\ \textbf{IX} $ is $\ \omega $ -continuous. Finally,
$\ max(\textbf{IX}) \simeq X $ where the set of maximal elements has the relative Scott topology from $\ \textbf{IX} $.

\noindent The observation  that the space-time has a canonical domain theoretic model,
teaches that from only a countable set of events and the causality relation,  space-time can be reconstructed in a purely order theoretic
manner using domain theory.

\noindent In [13], K. Martin and P. Panangaden  construct the space-time from a discrete causal set as follows:\\
An abstract basis is a set $\ (C, \ll) $ with a transitive relation that is interpolative
from the $\ - $ direction: $\ F \ll x \Rightarrow \exists y \in C \ni: F \ll y \ll x $ for all
finite subsets $\ F \subseteq C $ and all $\ x \in F $. Suppose, it is also interpolative from the $\ + $ direction:
$\ x \ll F  \Rightarrow \exists y \in C \ni: x \ll y \ll F $.
Then a new abstract basis of intervals can be defined as,
$\ int(C) = \{ (a, b) / a \ll b \} = \ll \subseteq C^{2} $ whose relation is
$\ (a, b) \ll (c, d) \equiv a \ll \ c \ and \ d \ll b$.

\noindent Let $\ \textbf{IC} $ denote the ideal completion of the abstract basis $\ int(C)$.

\noindent\textbf{Theorem 4.2.18:} Let $\ C $ be a countable dense subset of a globally hyperbolic
space-time $\ M $ and $\ \ll = I^{+} $ be timelike causality. Then $\ max(\textbf{IC})\simeq M $
where the set of maximal elements have the Scott topology.

\noindent This theorem is very different because, a process by which a countable set with a
causality relation determines a space, is identified here in abstract terms. The process is entirely order theoretic
in nature and space-time is not required to understand or execute it. In this sense, the understanding of the
relation between causality and the topology of space-time is  explainable
independently of geometry.

\noindent In a $\ C^{0} $- globally hyperbolic space-time, we can now extend
some of the order theoretic concepts to K- causality. To
generalize some of these concepts in the context of K- causality,
we first prove the following.

\noindent\textbf{Proposition 4.2.19:} \ In a $\ C^{0} $ - globally
hyperbolic space-times, $\ x  \ll y \ \Rightarrow \  y \in
K^{+}(x) $ where the partial order is $\ \prec = K^{+}   $.

\noindent\textbf{Proof :} \ Let $\ x  \ll y $. Consider $\ int
K^{-}(y) $ which is an open set not containing $\ y $. Since $\ y
\in K^{-}(y), \ y $ is a limit point of $\ int K^{-}(y)$. Hence
there exists a sequence $\ y_{n} $ in $\ int K^{-}(y)$ such that
lim $\ y_{n}  = y  $. We can choose $\ y_{n}$ as increasing
sequence. (using second countability as in Lemma 4.3 of [13]). Thus sup $\ y_{n} = y $. Now $\{ y_{n} \} $ is a
directed set with supremum $\ y $. Hence  $\ \ \exists \ y_{n} $
in $\ int K^{-}(y) $ such that $\ x \prec y_{n} \prec y $, as $\ x
\ll y $. Thus $\ y \in K^{+}(x) $.

\noindent It must be noted that above analysis does not require any kind of
differentiability conditions on a space-time manifold, and results
are purely topological and order- theoretic.

\noindent We also have, analogous to above,

\noindent\textbf{Definition 4.2.16:}\
$\ \Downarrow x = \{ a \in  M  /  a \ll x  \}$ and $\ \Uparrow x = \{ a \in  M  /  x \ll a \}$ .

\noindent Since $\ a \ll x \Rightarrow a \in K^{-}(x) $, we have, \\
$\ \Downarrow x \in K^{-}(x) $ and $\ \Uparrow x \in K^{+}(x) $.

\noindent We illustrate, for Lawson topology, as to how the concepts above
can be generalized to K- causality.

\noindent\textbf{Proposition 4.2.20:} \ Lawson topology, in K- sense, is
contained  in the manifold topology.

\noindent\textbf{Proof :} \ Let us  take a basis for Lawson
topology as the sets of the
form \\
$\{  \Uparrow x \sim \uparrow F  \ / \ F $ is finite $\}$. Since F
is finite, F is compact in the manifold topology and hence $\
\uparrow F $ is closed. Since the sets $\ \Downarrow x $ and $\
\Uparrow x $ are open in the manifold topology ( in a $\ C^{0} $ -
globally hyperbolic space-time ), $\ \Uparrow x \sim \uparrow F $
are also open in the manifold topology .

\noindent Thus Lawson open sets are open  in the manifold topology also and
hence the result follows.

\noindent Similar analysis can be given for Scott topology and interval
topology also. The intervals defined above, with appropriate cone
structure coincide with causal intervals and hence so does the definition of global hyperbolicity. When the
partial order is $\ J^{+} $, interval topology coincides with
Alexandrov topology and as is well-known, in a strongly causal
space-time, Alexandrov topology coincides with the manifold
topology.

\section{ Concluding Remarks}

We note that there are a large number of space-times (solutions of Einstein field equations) which are inhomogeneous
(see Krasinski [46]) and hence do not fall in the above class: $\ M = G/H $. M.Rainer [42] defines yet another partial order
using cones as subsets of a topological manifold and a differential manifold (space-time) which is a causal relation in
the sense defined above and which is more general than $\ J^{+}$. Rainer, furthermore defines analogous causal hierarchy like in the
classical causal structure theory. Of course, for Minkowski space, the old and new definitions coincide. For a $\ C^{2}$-globally
hyperbolic space-time $\ J^{+}, K^{+}$  and Rainer's relation all coincide, whereas for a $\ C^{0}$-globally hyperbolic space-time,
$\ K^{+}$ and Rainer's relation on topological manifold coincide.

Moreover if the cones are characteristic surfaces of the Lorentzian metric, then all his definitions of causal hierarchy coincide with the classical definitions. (cf theorem 2 of Rainer [42]). For more details on this partial order, we refer the reader to this paper.

B. Carter [47] discusses causal relations from a different perspective and discusses in detail many features of this relationship. Topological considerations in the light of time-ordering have been discussed by E.H.Kronheimer [48].

Using cone structure, a Causal Topology on Minkowski space was first discussed by Zeeman [49] way back in 1967. This topology has many interesting
features. At the same time, it is difficult to handle mathematically because it is not a normal topological space. Gheorghe and Mihul  [29] introduced another topology on Minkowski space by using causal relation and where it was assumed that a positive cone is closed in the Euclidean topology. They further proved that this topology coincides with Euclidean topology. R. Gobel [50] worked out in details many features of Zeeman like topologies in the context of space-time of general relativity. Around the same time, Hawking, King and McCarty [51] and Malament [52] worked out interesting features of a topology on space-time in general relativity using time-like curves. Though this work is mathematically interesting, it did not receive much response from people working in General Relativity. In 1992, D. Fullwood [53] constructed another causal topology $F$ from a basis of sets obtained by taking the union of two Alexandrov intervals $\ < x, y, z> \equiv < x, y > \cup < y, z > \cup \ y $. These sets are not open in the manifold topology since they include the intermediate point $y$. $F$ contains information about the space-time dimension and $F$ is Hausdorff iff the space-time is future and past distinguishing and is moreover, strictly finer than the manifold topology.

Fullwood showed that $F$ can also be obtained via a causal convergence criterion on time-like sequences of events. Recently, Onkar Parrikar and Sumati Surya [54] generalized this definition to include all monotonic causal sequences. This gives rise to yet another causal topology which is denoted by $P$. They showed that $P$ is strictly coarser than $F$ and also strictly finer than the manifold topology. The paper by Parrikar and Surya gives a non-trivial generalization of the MHKM (Malament-Hawking-King-McCarty) theorem and suggests that there is a causal topology for FPD (Future and Past Distinguishing) space-times which encodes manifold dimension and which is strictly finer than the Alexandrov topology. The construction uses a convergence criterion based on sequences of \emph{chain-intervals} which are the causal analogs of null geodesic segments. They also show that when the region of strong causality violation satisfies a local achronality condition, this topology is equivalent to the manifold topology in an FPD space-time. This work is motivated by Sorkin's Causal sets approach to Quantum Gravity. A somewhat different and more topological approach has been adopted by Martin Kovar [55]. For more details on Zeeman-like topologies and their relationship with manifold topology of space-time, we refer the reader to [56].

\section{References}

1. A. D. Alexandrov, On Lorentz transformation, \emph{Uspekhi. Math. Nauk.} \textbf{5 }(1950) 187 . \\
2. A. D. Alexandrov, Mappings of spaces with families of cones and space-time transformations. \emph{Annali di Matematica Pura ed Aplicata}, \textbf{103} (1967) 229-257 .\\
3. A. D. Alexandrov, A contribution to chronogeometry, \emph{Canadian Jour. of Math.}, \textbf{19} (1967)  1119-1128 .\\
4. S.W. Hawking and G.F.R. Ellis, The Large Scale Structure of Space-time ( Cambridge University Press, 1973).\\
5. R. Wald,  \emph{General Relativity} (University of Chicago Press, 1984).\\
6. P.S. Joshi, \emph{Global Aspects in Gravitation and Cosmology} (Oxford Science Publications, 1993).\\
7. F. Dowker, J. Henson and R. Sorkin, Quantum gravity phenomenology, Lorentz invariance and discreteness, \emph{Mod. Phys. Lett.}, \textbf{A19} (2004)  1829-1840.\\
8. R. Sorkin and E.Woolgar,  A Causal Order for Spacetimes with $C^0$ Lorentzian Metrics: Proof of Compactness of the Space of Causal Curves, \emph{Class. Quantum Grav}, \textbf{3} (1996) 1971-1994.\\
9. Sujatha Janardhan and R.V. Saraykar, K-causal structure of space-time in general  relativity, \emph{Pramana-Journal of Physics}, \textbf{70} (2008) 587-601.\\
10. Sujatha Janardhan and R.V. Saraykar, Domain theoretic approach to causal structure on a space-time manifold, \emph{International Jour. of Math. Sci. and Engg. Appls.} \textbf{4} (2010) 47-58.\\
11. E. Minguzzi, Chronological space-times without lightlike lines are stably causal, \emph{Comm. Math. Phys.} \textbf{288} (2009)  801-819.\\
12. E. Minguzzi, K-causality coincides with stable causality, \emph{Comm. Math. Phys.} \textbf{290} (2009)  239-248.\\
13. K. Martin and P. Panangaden, A domain of space-time intervals in general relativity, \emph{Commun. Math. Phys.} \textbf{267} (2006)  563-586.\\
14. V.A. Truong and L. Tuncel, Geometry of homogeneous cones: duality mapping and optimal self-concordant barriers, \emph{Math. Program. Ser.} \textbf{A. 100} (2004) 295-316.\\
15. R.D. Luce and R. Raifa, \emph{Games and Decisions: Introduction and Critical survey} ( Dover, New York, 1989).\\
16. V. Kreinovich and O. Kosheleva, From (Idealized) Exact Causality-Preserving  Transformations to Practically Useful Approximately-Preserving Ones: A General Approach,  \emph{Inter. Jour. Theo. Phys.} \textbf{47}(2008) 1083-1091.\\
17. V. Kreinovich and O. Kosheleva, Computational Complexity of Determining which Statements about Causality Hold in Different Space-Time Models, \emph{Theoretical Computer Science,} \textbf{405} (2008) 50-63.\\
18. E. Minguzzi and M. Sanchez, The causal hierarchy of spacetimes : in H. Baum and D. Alekseevsky (eds.), Recent developments in pseudo-Riemannian geometry, \emph{ESI Lect. Math. Phys.}, (Eur. Math. Soc. Publ. House, Zurich, 2008) (2008)  299-358.\\
19. J. Hilgert and G. Olafsson, \emph{Causal Symmetric Spaces-Geometry and Harmonic Analysis, Perspectives in Mathematics - Series,} Vol.18 ( Academic Press, 1997).\\
20. R. Gilmore, \emph{Lie Groups, Physics, and Geometry} ( Cambridge University Press, 2008).\\
 21. J.A. Wolf, \emph{Spaces of Constant Curvature} (McGraw-Hill Book Company, 1967).\\
22.  A. Knapp, \emph{Lie groups beyond an introduction. Series: Progress in Mathematics}, Vol. 140 (Birkhausar, Basel, 2002).\\
23. L. Fuchs, \emph{Partially Ordered Algebraic Systems} (Pergamon Press, 1963).\\
24. J. Hilgert and K.H. Neeb, Groupoid C*-algebras of order compactified symmetric spaces, \emph{Japnese Jour. Math.} \textbf{21} (1995) 117-188.\\
25. I. Chajda and S. Hoskava, A characterization of cone preserving mappings of quasi-ordered sets, \emph{Miskolc Mathematical notes}, \textbf{6} (2005) 147-152.\\
26. J.A. Lester and M.A. McKiernan, On null cone preserving mappings, \emph{Math. Proc. Cambridge Phil. Soc}. \textbf{81}(1977) 455-462.\\
27. J.A. Lester, A physical characterization of conformal transformations of Minkowski space-time, \emph{Annals of Discrete Mathematics},\textbf{18 } (1983) 567-574.\\
28. F. Zapata and V. Kreinovich, Reconstructing an Open Order from Its Closure, with Applications to Space-Time Physics and to Logic, \emph{Studia logica}, \textbf{87} (2011) 1-15.\\
29. C. Gheorghe and E. Mihul, Causal groups of space-time, \emph{Commun. Math. Phys.} \textbf{14 }(1969)  165-170.\\
30. R. Penrose, \emph{Techniques of Differential Topology in Relativity } (AMS Colloquium, SIAM Publications, 1972).\\
31. A. Garcia-Parrado and J.M. Senovilla, General study and basic properties of causal symmetries, \emph{Class. Quantum Grav.} \textbf{21}(2004) 661-696.\\
32. A. Garcia-Parrado and J.M. Senovilla, Causal Relationship: a new tool for the causal characterization of Lorentzian manifolds, \emph{Class. Quantum Grav.} \textbf{20} (2003) 625-664.\\
33. S.W. Hawking and R.K. Sachs, Causally continuous space-times, \emph{Commun. Math. Phys.} \textbf{35} (1974) 287-296.\\
34. F. Dowker, R.S. Garcia and Sumati Surya, Morse Index and Causal Continuity: A Criterion for Topology Change in Quantum Gravity, \emph{Class. Quant. Grav}. \textbf{17} (2000)  697-712.\\
35. M. Sanchez, Causal hierarchy of space-times, temporal functions and smoothness of Geroch splitting. A revision.  \emph{Matematica Contemporanea},  \textbf{29} (2005)  127-155.\\
36. H.J. Seifert, The causal boundary of space-times,  \emph{Gen. Rel. And Grav.} \textbf{1} (1971)  247-259.\\
37. N.M.J. Woodhouse, The differential and causal structures of space-time, \emph{Jour. Math. Phys.}, \textbf{14} (1973)  495-501.\\
38. E. Minguzzi, E. The causal ladder and the strength of K-causality. I. \emph{Class. Quantum Grav.} \textbf{25} (2008)  015009.\\
39. E. Minguzzi, The causal ladder and the strength of K-causality. II. \emph{Class. Quantum Grav.} \textbf{25} (2008)  015010.\\
40. E. C. Zeeman, Causality implies Lorentz group,  \emph{Jour. Math. Phys.} \textbf{5} (1964) 490-493.\\
41. H.J. Borchers and G.C. Hegerfeldt,  The Structure of Space-time Transformations, \emph{Commun. Math. Phys.} \textbf{28} (1972)  259-266.\\
42. M. Rainer, Cones and causal structures on topological and differentiable manifolds. \emph{Jour.Math.Phys.} \textbf{40} (1999) 6589-6597. Erratum - ibid, (2000) 41, 3303.\\
43. S. Abramsky and A. Jung,  \emph{Domain Theory}, In: S.Abramsky, D. M.Gabbay, T. S. E.Maibaum, Editors, Handbook of Logic in Computer Science, Vol. III. (Oxford University Press 1994).\\
44. K. Martin and P. Panangaden, New Structures for Physics. \emph{Lecture Notes in Physics,}  \textbf{813} (2011)  687-703.\\
45. R. Penrose, R. Sorkin and E. Woolgar, A Positive Mass Theorem Based on the Focusing and Retardation of Null Geodesics. (1993)  arXiv: gr-qc/9301015.\\
46. A. Krasinski, \emph{Inhomogeneous Cosmological Models.}(Cambridge University Press, New Ed. 2006).\\
47. B. Carter, Causal structure in space-time. \emph{Gen. Rel. Grav.} \textbf{1} (1971) 349-391.\\
48. E.H. Kronheimer, Time Ordering and Topology, \emph{Gen. Rel. Grav.} \textbf{1}(1971)  261- 268.\\
49. E.C. Zeeman, The topology of Minkowski space, \emph{Topology,} \textbf{6} (1967)  161 - 170.\\
50. R. Gobel, Zeeman Topologies on Space-Times of General Relativity Theory, \emph{Commun. math. Phys.} \textbf{46} (1976)  289-307.\\
51. S.W. Hawking, A.R. King and P.J. McCarthy, A New Topology for curved space-time which incorporates the causal, differential and conformal structures, \emph{Jour. Math. Phys.} \textbf{17}(1976) 174-181.\\
52. B. Malament, The class of continuous time-like curves determines the Topology  of space-time, \emph{Jour. Math. Phys.}\textbf{18} (1977) 1399-1404.\\
53. D. Fullwood, A new topology of space-time.  \emph{Jour. Math. Phys.} \textbf{33} (1992) 2232-2241.\\
54. O. Parrikar and Sumati Surya, Causal Topology in Future and Past Distinguishing Spacetimes, \emph{Class.Quant.Grav.} \textbf{28} (2011) 155020.\\
55. M. Kovar, A new causal topology and why the universe is co-compact, (2011) arXiv: gr-qc/1112.0817v2.\\
56. R.V. Saraykar and Sujatha Janardhan, Zeeman-like topologies in special and general theory of relativity, (2014) arXiv: gr-qc/ 1410.3895.

\end{document}